\documentclass[
    aps,
    physrev,
]{revtex4-2}

\usepackage[colorlinks=true, allcolors=blue]{hyperref}
\usepackage{mathtools, amsfonts, bm}
\usepackage{graphicx}

\usepackage{amsthm}
\theoremstyle{plain}
\newtheorem{theorem}{Theorem}[section]
\newtheorem{lemma}[theorem]{Lemma}

\DeclareMathOperator*{\argmin}{arg\;min}

\DeclareMathOperator*{\sign}{sign}

\begin{document}

\title{
    Phase transition in compressed sensing using \\ log-sum penalty and adaptive smoothing
}

\author{Keisuke Morita}
\email[]{mory22k@dc.tohoku.ac.jp}
\affiliation{Graduate School of Information Sciences, Tohoku University, Sendai 980-8579, Japan}

\author{Federico Ricci-Tersenghi}
\affiliation{Department of Physics, Sapienza University of Rome, Rome 00185, Italy}

\author{Masayuki Ohzeki}
\affiliation{Graduate School of Information Sciences, Tohoku University, Sendai 980-8579, Japan}
\affiliation{Department of Physics, Institute of Science Tokyo, Tokyo 152-8550, Japan}
\affiliation{Research and Education Institute for Semiconductors and Informatics, Kumamoto University, Kumamoto 860-8555, Japan}
\affiliation{Sigma-i Co., Ltd., Tokyo 108-0075, Japan}

\date{\today}

\begin{abstract}
    In many real-world problems, recovering sparse signals from underdetermined linear systems remains a fundamental challenge.
    Although $\ell_1$ norm minimization is widely used, it suffers from estimation bias that prevents it from reaching the Bayes-optimal reconstruction limit.
    Nonconvex alternatives, such as the log-sum penalty, have been proposed to promote stronger sparsity.
    However, maintaining their algorithmic stability is challenging.
    To address this challenge, we introduce an adaptive smoothing strategy within an approximate message passing framework to mitigate algorithmic instability.
    Furthermore, we evaluate the typical exact-recovery threshold for Gaussian measurement matrices using the replica method and state evolution.
    The results indicate that the adaptive method achieves exact recovery over a broader region than $\ell_1$-norm minimization, although metastable states hinder the attainment of the information-theoretic limit.
\end{abstract}

\maketitle

\tableofcontents

\section{Introduction}
\label{sec:introduction}

Compressed sensing is a framework for recovering sparse signals from fewer measurements than the signal dimension.
Its basic idea is that, although a signal may be high-dimensional, the amount of essential information it contains is often much smaller because many of its components are zero or negligibly small.
This property enables reconstruction of the original signal from incomplete observations under suitable conditions, and it has found many real-world applications in areas such as accelerated magnetic resonance imaging (MRI) \cite{Lustig2007-zw}, computational imaging \cite{Duarte2008-sm, Wagadarikar2008-ef}, and millimeter-wave communications \cite{Ayach2014-qg}.

In the standard setting, compressed sensing is formulated as an underdetermined linear system, where the number of measurements is smaller than the signal dimension.
The goal is to identify the sparsest signal among all solutions consistent with the observations.
In principle, this can be achieved by minimizing the $\ell_0$ pseudo-norm, which counts the number of nonzero components.
However, $\ell_0$ minimization is nonconvex and NP-hard, and is therefore computationally intractable in general.
For this reason, $\ell_1$ norm minimization, a convex relaxation of $\ell_0$ minimization, has become the standard approach.

The reconstruction performance of these specific formulations can be characterized from two different perspectives:~theoretical guarantee and typical performance.
The theoretical guarantee provides mathematical conditions under which exact reconstruction is guaranteed for worst-case problem instances, such as conditions based on the restricted isometry property (RIP) \cite{Candes2008-de} and the null space property \cite{Foucart2013-ej}.
Meanwhile, the typical performance characterizes the reconstruction threshold for typical problem instances in the high-dimensional limit under a random measurement matrix and a stochastic signal model.

Two complementary frameworks are used to characterize typical performance.
One framework evaluates typical algorithmic performance using approximate message passing (AMP) together with state evolution (SE).
AMP provides an explicit iterative reconstruction algorithm~\cite{Donoho2009-jl,Donoho2010-mz}, while SE predicts its asymptotic behavior in the high-dimensional limit~\cite{Donoho2010-kk,Bayati2011-de}.
Another framework characterizes the typical reconstruction threshold using the replica method, a statistical-mechanical approach based on free-energy analysis~\cite{Kabashima2009-gx,Ganguli2010-ef}.
In representative settings such as $\ell_1$ norm minimization, the SE prediction and the typical reconstruction threshold characterized by the replica method yield consistent predictions.

To quantitatively evaluate this method-dependent typical performance, one requires a theoretical benchmark that represents the best performance permitted by the statistical structure of the problem.
Such a benchmark should be independent of any specific reconstruction method.
This role is played by the Bayes-optimal setting, the ideal setting where the true signal prior and observation model are correctly specified, so that all information available for reconstruction is fully exploited~\cite{Krzakala2012-ot}.
From this setting, two fundamental thresholds regarding the ratio of the number of observations to the number of nonzero components in the true signal are identified:~the information-theoretic threshold (IT threshold) and the Bayes-optimal spinodal threshold (BO threshold).
The IT threshold is the threshold above which the exact-recovery state can exist as a stable fixed point.
When the ratio is smaller than this threshold, the exact-recovery state is unstable, and an algorithm cannot remain at the exact recovery state even if it is initialized there.
The BO threshold, on the other hand, is the threshold above which exact recovery is the \textit{unique} stable fixed point.
Below this threshold, a metastable state that corresponds to finite reconstruction error emerges.
Since the algorithm converges to this metastable state unless it is initialized in a sufficiently close neighborhood of the exact-recovery state, it cannot achieve exact recovery.
The BO threshold, therefore, provides a fundamental benchmark for the reconstruction performance that is algorithmically achievable.

From this perspective, although $\ell_1$ norm minimization has theoretical guarantees~\cite{Donoho2006-jx,Candes2006Stable} and computational tractability~\cite{Donoho2009-jl,Donoho2010-kk,Donoho2010-mz}, it also has intrinsic limitations.
Because the $\ell_1$ penalty imposes nearly uniform shrinkage regardless of coefficient magnitude, even large true coefficients are biased toward zero~\cite{Zou2006-qw}.
It may also produce false positives, in which truly zero components are estimated as small but nonzero~\cite{Fan2001-pc}.
More fundamentally, its reconstruction threshold generally falls short of the BO threshold~\cite{Kabashima2009-gx,Ganguli2010-ef,Krzakala2012-is}.

To overcome the limitations of the $\ell_1$ penalty, nonconvex penalty functions have emerged as a strategy to induce stronger sparsity and to mitigate estimation bias.
Examples include \textit{smoothly clipped absolute deviation} (SCAD) \cite{Fan2001-pc}, $\ell_p$ quasi-norm with $0 < p < 1$ \cite{Chartrand2007-so}, \textit{minimax concave penalty} (MCP) \cite{Zhang2010-gb}, and log-sum penalty \cite{Coifman1992-mi, Candes2008-de}.
A common feature of these nonconvex penalties is strong penalization near the origin and progressively weaker marginal penalization for large signals.
This feature promotes stronger sparsity while reducing overshrinking bias.
To exploit these properties of nonconvex penalties, specific algorithms such as iteratively reweighted least squares (IRLS) \cite{Daubechies2010-sj, Chartrand2008-rl}, and AMP \cite{Donoho2009-jl, Sakata2018-ey, Gu2025-wr} have also been proposed.

While the introduction of nonconvex penalties offers significant theoretical advantages, it simultaneously presents algorithmic challenges.
In contrast to $\ell_1$ minimization, signal reconstruction using a nonconvex penalty can encounter algorithmic issues, including sensitivity to initialization and difficulty escaping local minima.
In addition, as the effective nonconvexity increases, the basin of attraction to fixed points generally shrinks, making algorithms more prone to divergence \cite{Sakata2021-du}.

These considerations motivate a focused study of a representative nonconvex penalty that enables the analysis of both recovery performance and algorithmic behavior.
Among various nonconvex functions, this study focuses on the log-sum penalty.

The log-sum penalty exhibits several properties in common with SCAD and MCP, which makes it suitable for both algorithmic and theoretical investigation.
Similar to SCAD and MCP, the log-sum penalty is a weakly convex function. This property enables the proximal operator to be treated as a continuous function within a specific parameter range \cite{Prater-Bennette2022-rz}.
This property allows the log-sum penalty to be incorporated into iterative algorithms such as the iterative shrinkage-thresholding algorithm (ISTA) \cite{Daubechies2004-tx,Beck2009-kk}, the alternating direction method of multipliers (ADMM) \cite{Boyd2010-hi}, and AMP.
On the theoretical side, the log-sum penalty has been proven to enable exact signal recovery under milder conditions on the RIP than the $\ell_1$ norm~\cite{Shen2013-ed}.
These properties place the log-sum penalty alongside other established nonconvex sparse penalties as a natural candidate for studying the relation between sparsity promotion and algorithmic stability.

The log-sum penalty also has several structural features that are particularly advantageous for the problem under consideration.
The degree of nonconvexity can be controlled through a single smoothing parameter, denoted by $\varepsilon$ throughout this paper.
This feature provides a simple continuation from a regime in which the proximal objective is convex to one that is more strongly nonconvex and promotes sparsity.
In contrast, SCAD and MCP require separate parameters to control the penalization scale (denoted by $\lambda$ in \cite{Fan2001-pc, Zhang2010-gb}) and the degree of concavity (denoted by $a$ in \cite{Fan2001-pc} and $\gamma$ in \cite{Zhang2010-gb}).
In addition, the curvature of the log-sum penalty varies continuously with the smoothing parameter.
Such continuous variation allows the sensitivity of the proximal operator and the stability of the associated AMP dynamics to be tracked as the nonconvexity increases.
Moreover, the logarithmic tail is naturally connected with heavy-tailed Bayesian shrinkage.
The logarithmically growing penalty is inherently linked to priors with polynomially decaying tails \cite{Bhadra2019-il}.
These properties make the log-sum penalty particularly suitable for investigating adaptive control of nonconvexity in AMP.

Despite these theoretical and computational properties, the typical reconstruction performance of the log-sum penalty under random measurements remains largely unexplored.
In particular, its typical exact-recovery threshold, the associated phase structure, and the relation between increasing nonconvexity and algorithmic instability remain unclear.

In this paper, we bridge this gap between theoretical guarantees and typical performance.
Our contributions are as follows.
We derive an AMP algorithm specialized to the log-sum penalty using the closed-form proximal operator and demonstrate the algorithmic instability arising from the penalty's nonconvexity.
To stabilize the algorithm, we propose an adaptive schedule for the smoothness parameter that keeps the proximal operator in the convex regime throughout the iterations.
In addition, we characterize the typical exact-recovery threshold for i.i.d.~Gaussian measurement matrices via SE and replica analysis, and demonstrate that the log-sum penalty with the proposed adaptive smoothing achieves exact recovery over a broader region than the $\ell_1$ penalty.

The rest of this paper is organized as follows.
In Sec.~\ref{sec:problem-setting}, we present the problem setting and describe the introduction of the log-sum penalty and the closed-form definition of the proximal operator.
In Sec.~\ref{sec:amp}, we explain the details of the AMP algorithm for the log-sum penalty and analyze its asymptotic behavior using SE.
In Sec.~\ref{sec:replica-analysis}, we use the replica method to analyze the macroscopic behavior of the system.
In Sec.~\ref{sec:phase-diagram}, we present the comprehensive phase diagram and discuss the relationship between the asymptotic behavior of the proposed method and reconstruction thresholds derived in Sec.~\ref{sec:replica-analysis}.
Finally, in Sec.~\ref{sec:discussion}, we discuss the results of this study and present future prospects.

\section{Problem setting and formulation}
\label{sec:problem-setting}

\subsection{Problem formulation}

Consider the reconstruction of an unknown original signal vector $\bm{x}^0 \in \mathbb{R}^N$ from a measurement matrix $\bm{A} \in \mathbb{R}^{M \times N}$ and an observed vector $\bm{y} \in \mathbb{R}^M$.
$\bm{y}$ is obtained by noise-free linear observation $\bm{y} = \bm{A}\bm{x}^0$.
The measurement rate, defined as the ratio of the number of measurements to the signal dimension, is denoted by $\alpha = M/N$.
In particular, we consider the underdetermined case where $0 < \alpha < 1$.
Each entry $A_{\mu i}$ of the measurement matrix $\bm{A}$ is i.i.d.~distributed according to $A_{\mu i} \sim \mathcal{N}(0, 1/N)$.
The original signal $\bm{x}^0$ is sparse, and each component $x^0_i$ follows the zero-mean Bernoulli--Gaussian distribution with sparsity level $\rho \in (0, 1)$:
\begin{align}
P_\text{true}(\bm{x}^0) = \prod_{i=1}^N \left( (1 - \rho) \delta(x_i^0) + \rho \phi(x_i^0) \right).
\label{eq:true-signal-distribution}
\end{align}
Here, $\delta(\cdot)$ is the Dirac delta function, and $\phi(\cdot)$ is the probability density function of the standard normal distribution $\mathcal{N}(0,1)$.
The expected number of nonzeros in the signal is $K = \rho N$.

The reconstruction of the sparse signal $\bm{x}^0$ using the log-sum penalty is formulated as
\begin{align}
    \hat{\bm{x}} = \argmin_{\bm{x}} R(\bm{x}) \quad \text{s.t.} \quad
    \bm{y} = \bm{A}\bm{x},
\label{eq:logsum-optimization}
\end{align}
where
\begin{align}
    R(\bm{x}) = \sum_{i=1}^N \log(|x_i| + \varepsilon).
\end{align}
Here, $\varepsilon > 0$ is a parameter. It controls the tradeoff between sparsity promotion and computational stability.

To further understand this optimization problem, it can be interpreted as the low-temperature limit of a related probabilistic system. Consider a Bayesian linear model with Gaussian noise of variance $\sigma^2$. The posterior probability distribution of $\bm{x}$ given $\bm{y}$ and $\bm{A}$ is
\begin{align}
    P_\beta(\bm{x} \mid \bm{y}, \bm{A}) \propto
    P_{\beta}(\bm{x}) \exp\!\left( - \frac{1}{2 \sigma^2} \| \bm{y} - \bm{A}\bm{x} \|_2^2 \right),
    \label{eq:posterior-distribution-logsum}
\end{align}
where $P_{\beta}(\bm{x})$ is the prior distribution parametrized by the inverse temperature $\beta$.
The prior distribution $P_{\beta}(\bm{x})$ is defined based on the penalty function as
\begin{align}
    P_{\beta}(\bm{x}) \propto \exp\!\left( - \beta R(\bm{x}) \right).
\end{align}
We can relate the constrained optimization to a Bayesian model via two distinct limits.
First, taking $\sigma^2 \to 0^+$ enforces the constraint $\bm{y}=\bm{A}\bm{x}$.
Second, taking $\beta \to \infty$ yields the global minimizer by concentrating the probability mass on the minimizers of $R(\bm{x})$ under the constraint.
By analyzing these limits, we can characterize the behavior of solutions to the optimization problem through the statistical properties of the corresponding posterior probability distribution.

The log-sum penalty function exhibits properties intermediate between the $\ell_1$ norm and the $\ell_0$ pseudo-norm.
When $\varepsilon \gg |x|$, the log-sum penalty can be approximated as $\log(|x|+\varepsilon) = \log(\varepsilon) + |x| / \varepsilon - O(|x|^2 / \varepsilon^2)$.
This approximation reveals that the penalty behaves like a scaled $\ell_1$ norm in the $\varepsilon \to \infty$ limit, satisfying
\begin{align}
\lim_{\varepsilon \to \infty} \varepsilon(R(\bm{x}) - N\log\varepsilon) = \|\bm{x}\|_1.
\end{align}
When $\varepsilon \ll |x|$, the log-sum penalty approaches the scaled $\ell_0$ norm.
For a fixed $\bm{x}$ with bounded nonzero components, in the infinitesimal $\varepsilon$ limit, we obtain
\begin{align}
    \lim_{\varepsilon \to 0^+} \frac{R(\bm{x}) - N\log\varepsilon}{\log(1+1/\varepsilon)} = \|\bm{x}\|_0.
\end{align}
Thus, these properties indicate that the smoothing parameter $\varepsilon$ governs the tradeoff between theoretical performance and algorithmic stability.
For instance, choosing $\varepsilon \gg |x|$ promotes global optimization by making the problem essentially convex.
However, this choice reduces solution sparsity and increases the number of required observations for exact reconstruction, as shown in Ref.~\cite{Prater-Bennette2022-rz}.
On the other hand, using $\varepsilon \ll |x|$ enhances sparsity of the solution but makes global optimization more difficult, turning the task into a challenging nonconvex problem.
Given these observations, the main point now is to consider how small $\varepsilon$ can be set in practice.
The key to answering this question lies in the specific behavior of the proximal operator.

\subsection{Proximal operator}

The behavior of sparse estimators can be described by the proximal operator.
The proximal operator is an important tool used in many iterative algorithms for sparse estimation, such as the AMP algorithm.
Given a vector $\bm{x} \in \mathbb{R}^N$, a penalty function $R: \mathbb{R}^N \to \mathbb{R}$, and a threshold level $\lambda \ge 0$, the proximal operator $\operatorname*{prox}_{\lambda R(\cdot)}$ is defined as the minimizer $\bm{z}$ of the linear combination of the squared $\ell_2$ norm of $\bm{x} - \bm{z}$ and $R(\bm{z})$.
In the remainder of this paper, we denote this operator as a function of $\bm{x}$ by $S(\bm{x}; \lambda R(\cdot)) = \operatorname*{prox}_{\lambda R(\cdot)}(\bm{x})$ and call it the \textit{thresholding function}.
That is, we define
\begin{align}
    S(\bm{x}; \lambda R(\cdot))
    = \operatorname*{prox}_{\lambda R(\cdot)}(\bm{x})
    = \argmin_{\bm{z}\in\mathbb{R}^N} \!\left\{ \frac{1}{2}\|\bm{z}-\bm{x}\|_2^2 + \lambda R(\bm{z}) \right\}.
\end{align}
Since the log-sum penalty term $R(\bm{z})$ is separable component-wise, $S(\bm{x}; \lambda R(\cdot))$ can also be computed component-wise. When the argument is a scalar $x$, $S(x; \lambda R(\cdot))$ represents a univariate thresholding function, while when it is a vector $\bm{x}$, $S(\bm{x}; \lambda R(\cdot))$ is a vector formed by applying the single-variable thresholding function to each part.

Here, we briefly review the closed form of the thresholding function derived in Ref.~\cite{Prater-Bennette2022-rz}.
The solution is obtained by minimizing the objective function
\begin{align}
    \varphi_{\lambda, x}(z) = \frac{1}{2}(z-x)^2 + \lambda \log\!\left( |z| + \varepsilon \right).
\end{align}
Depending on the values of the regularization parameter $\lambda$ and smoothness parameter $\varepsilon$, we have two situations:~
the convex case $\sqrt{\lambda} \le \varepsilon$ and the nonconvex case $\sqrt{\lambda} > \varepsilon$.

In the convex case $\sqrt{\lambda} \le \varepsilon$, the objective function $\varphi_{\lambda, x}(z)$ has a unique local minimum, and $S$ is a continuous function with respect to the input. The thresholding function in this case is given by
\begin{align}
    S(x; \lambda R(\cdot)) =
    \begin{dcases}
        \sign(x) r_+(|x|) & \text{if} ~ |x| > \frac{\lambda}{\varepsilon}, \\
        0 & \text{otherwise},
    \end{dcases}
    \label{eq:proximal-operator-logsum-convex}
\end{align}
where $r_+(x)$ is the positive root of $\varphi_{\lambda, x}'(z)=0$ for $x > 0$, given by
\begin{align}
    r_+(x) = \frac{x-\varepsilon}{2} + \sqrt{\frac{(x+\varepsilon)^2}{4} - \lambda}.
\end{align}

In the nonconvex case $\sqrt{\lambda} > \varepsilon$, on the other hand, $\varphi_{\lambda, x}(z)$ has two local minima:~$z\in\{0, \sign(x) r_+(|x|)\}$.
The solution switches between these two options at a threshold $|x| = x_c$, which is the unique root of $c(x) = \varphi_{\lambda, x}(r_+(x)) - \varphi_{\lambda, x}(0)$ on the interval $x \in [2\sqrt{\lambda}-\varepsilon, \lambda/\varepsilon]$.
This switching of the global minimum $z$ makes $S$ discontinuous.
The closed-form solution is given as
\begin{align}
    S(x; \lambda R(\cdot)) =
    \begin{dcases}
        0
            & \text{if } |x| < x_c, \\
        0 \text{ or } \sign(x) r_+(|x|)
            & \text{if } |x| = x_c, \\
        \sign(x) r_+(|x|)
            & \text{if } |x| > x_c.
    \end{dcases}
    \label{eq:proximal-operator-logsum-nonconvex}
\end{align}
Note that for $|x| = x_c$, both $z=0$ and $z=\sign(x) r_+(|x|)$ are global minimizers of $\varphi_{\lambda, x}(z)$.
In subsequent numerical experiments, we select $z = 0$.
This choice does not affect the asymptotic analysis and has a negligible impact on numerical computations.
The threshold $x_c$ can be found numerically, for example, by bisection search.
Here, we simply compute $\varphi_{\lambda, x}(z)$ for both candidates $z\in\{0, \sign(x) r_+(|x|)\}$ and use the one with the smaller objective value.

At points where the thresholding function is differentiable, implicit differentiation of the stationarity condition gives
\begin{align}
    S'(x; \lambda R(\cdot)) =
    \begin{dcases}
        \dfrac{(|z| + \varepsilon)^2}{(|z| + \varepsilon)^2 - \lambda}
        & \text{if } |x| > x_\text{sw}, \\
        0
        & \text{if } |x| < x_\text{sw},
    \end{dcases}
    \label{eq:logsum-derivative}
\end{align}
where $z = S(x; \lambda R(\cdot))$.
The switching points are given as $x_\text{sw} = \lambda/\varepsilon$ for the convex regime $\varepsilon \ge \sqrt{\lambda}$ and $x_\text{sw} = x_c$ for the nonconvex regime $\varepsilon < \sqrt{\lambda}$.

\begin{figure}[tbp]
    \centering
    \includegraphics[width=0.9\columnwidth]{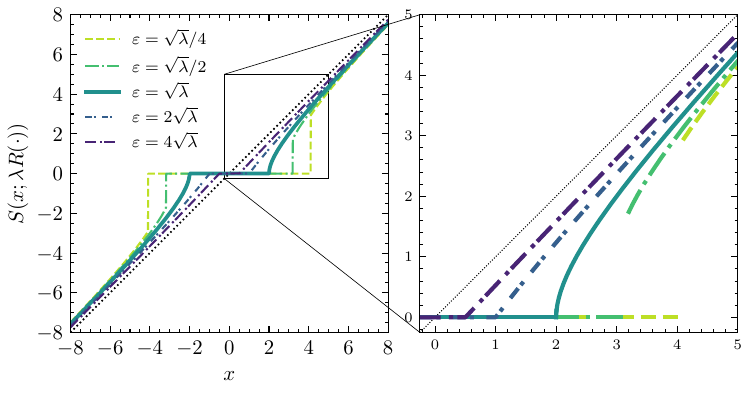}
    \caption{
        Thresholding function $S(x; \lambda R(\cdot))$ for the log-sum penalty with $\varepsilon=\sqrt{\lambda}/4$, $\sqrt{\lambda}/2$, $\sqrt{\lambda}$, $2 \sqrt{\lambda}$, and $4 \sqrt{\lambda}$ with $\lambda=4$.
        The thick solid line representing $\varepsilon = \sqrt{\lambda}$ indicates the limit at which the threshold function remains continuous.
        }
    \label{fig:proximal-operator}
\end{figure}

Figure \ref{fig:proximal-operator} shows the thresholding function $S(x; \lambda R(\cdot))$ in both regimes of $\varepsilon$.
In the convex regime $\varepsilon \ge \sqrt{\lambda}$, $S$ is continuous. As $\varepsilon \to \infty$, $S$ approaches the soft-thresholding function used in $\ell_1$ minimization.
Conversely, when $\varepsilon$ is in the nonconvex regime $\varepsilon < \sqrt{\lambda}$, the thresholding function displays \textit{jumps} at $x = \pm x_c$. As $\varepsilon \to 0^+$, its behavior increasingly resembles that for $\ell_0$ minimization.

The discontinuity of the thresholding function in the nonconvex case is critical for the stability of algorithms.
In terms of algorithmic behavior, these jumps mean that small noise or perturbations in the input are amplified into large disturbances in the output.
To achieve stable signal reconstruction, we need to control the parameter $\varepsilon$ to ensure the thresholding function remains continuous.
The next section gives a concrete example of this adaptive control.

\section{Approximate message passing and adaptive smoothing}
\label{sec:amp}

\subsection{Approximate message passing for log-sum penalty}

This section formulates the update rules of the AMP algorithm for the compressed sensing problem addressed in this study.
AMP is an iterative algorithm that computes marginal probabilities, a key step in solving inference problems in high dimensions \cite{Donoho2009-jl}.
The update rules of the AMP for the minimization of the log-sum penalty with i.i.d.\ zero-mean Gaussian measurements are
\begin{align}
    h_{i}^{[t]}
        &= \hat{x}_{i}^{[t]} + \frac{1}{\alpha} \sum_{\mu=1}^M A_{\mu i} z_\mu^{[t]}, \label{eq:amp-start} \\
    \hat{x}_{i}^{[t+1]}
        &= S\!\left(h_{i}^{[t]}; \lambda^{[t]} R(\cdot)\right), \\
    z_\mu^{[t+1]}
        &= y_\mu - \sum_{i=1}^N A_{\mu i} \hat{x}_{i}^{[t+1]} + \frac{z_\mu^{[t]}}{\alpha} \frac{1}{N} \sum_{i=1}^N S'\!\left(h_{i}^{[t]}; \lambda^{[t]} R(\cdot)\right), \label{eq:amp-z_mu-with-onsager}
\end{align}
where the threshold level $\lambda^{[t]}$ is updated as:
\begin{align}
    \chi^{[t+1]}
        &= \lambda^{[t]} \frac{1}{N} \sum_{i=1}^N S'\!\left(h_{i}^{[t]}; \lambda^{[t]} R(\cdot)\right), \\
    \lambda^{[t+1]} &= \frac{\chi^{[t+1]}}{\alpha}. \label{eq:amp-end}
\end{align}
Here, at iteration $t$, $h_i^{[t]}$ is an effective observation, $\hat{x}_i^{[t]}$ is the estimated signal, $z_\mu^{[t]}$ is a measurement residual with Onsager correction, and $\chi^{[t]}$ is the susceptibility; $i \in \{1,\dots,N\}$ is the signal index, and $\mu \in \{1,\dots,M\}$ is the measurement index.
Note that $\lambda^{[t]}$ should not be confused with either the parameter $\lambda$ in Refs.~\cite{Fan2001-pc, Zhang2010-gb} or the inverse temperature $\beta$.
The update rule of $\lambda^{[t]}$ is obtained by approximating belief propagation in the case of $N \gg 1$ using the same approach as in Ref.~\cite{Donoho2010-mz}.

The introduction of the log-sum penalty increases the computational cost only by a constant factor.
For the $\ell_1$ penalty, the soft-thresholding function can be evaluated in $O(1)$ operations per component, and thus the time complexity per AMP iteration is dominated by the matrix-vector multiplications and is $O(MN)$.
The thresholding function for the log-sum penalty also requires only $O(1)$ operations per component, although its evaluation involves a larger constant number of scalar operations.
Therefore, replacing the $\ell_1$ soft-thresholding function with the log-sum thresholding function does not change the asymptotic per-iteration computational complexity.

For i.i.d.\ Gaussian measurement matrices and suitable regularity conditions on the scalar update functions, the asymptotic dynamics of AMP are characterized by SE \cite{Bayati2011-de, Krzakala2012-ot}.
The third term in the right-hand side of Eq.~\eqref{eq:amp-z_mu-with-onsager} cancels the leading-order self-interaction generated by reusing the same measurement matrix $\bm{A}$ across iterations.
This cancellation yields the asymptotically Gaussian scalar effective observation described by SE.
SE tracks the time evolution of the mean squared error (MSE), which is defined by
\begin{align}
    \text{MSE}^{[t]} = \frac{1}{N}\|\hat{\bm{x}}^{[t]} - \bm{x}^0\|_2^2,
\end{align}
in the $N \to \infty$ limit.
Here, $\hat{\bm{x}}^{[t]} \in \mathbb{R}^N$ is the estimated signal at iteration $t$ and $\bm{x}^0 \in \mathbb{R}^N$ is the true signal.
The SE equations for the aforementioned AMP algorithm are given as
\begin{align}
    h^{[t]} &= x^0 + \sqrt{\frac{\text{MSE}^{[t]}}{\alpha}} \xi,
    \label{eq:SE-effective-field} \\
    \text{MSE}^{[t+1]} &= \mathbb{E}_{x^0} \int D \xi\, \left( S \!\left( h^{[t]}; \frac{\chi^{[t]}}{\alpha} R(\cdot) \right) - x^0 \right)^2,
    \label{eq:SE-mse} \\
    \chi^{[t+1]} &= \frac{\chi^{[t]}}{\alpha} \mathbb{E}_{x^0} \int D \xi\, S' \!\left( h^{[t]}; \frac{\chi^{[t]}}{\alpha} R(\cdot) \right),
    \label{eq:SE-chi}
\end{align}
where $D \xi = \phi(\xi) d\xi$ is the standard Gaussian measure.

\begin{figure*}[tbp]
    \centering
    \includegraphics[width=1.0\textwidth]{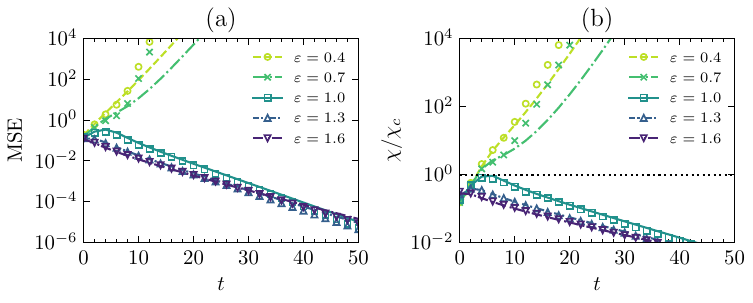}
    \caption{
        Time evolution of (a) MSE and (b) $\chi/\chi_c$ where $\chi_c = \varepsilon^2 \alpha$ when $\varepsilon$ is fixed to various values.
        In panel (b), $\chi/\chi_c > 1$ indicates that the thresholding function is discontinuous.
        The symbols represent the numerical results by AMP ($N=10^4$), while the curves indicate the theoretical predictions by SE ($N \to \infty$).
        Measurement rate $\alpha=0.5$ and signal density $\rho=0.2$.
    }
    \label{fig:amp-fixed}
\end{figure*}

Numerical experiments demonstrate that the stability of AMP depends on the parameter $\varepsilon$, as shown in Fig.~\ref{fig:amp-fixed}.
Figure \ref{fig:amp-fixed}(a) shows the time evolution of MSE by AMP and SE when $\varepsilon$ is fixed to various values.
Here, we initialize $\hat{\bm{x}}^{t=0}=\bm{0}$ and $\chi^{t=0}=1$ for AMP and $\text{MSE}^{t=0} = \rho$ and $\chi^{t=0}=1$ for SE.
In the case of $\varepsilon \ge 1$, MSE monotonically decreases and converges to zero.
However, when $\varepsilon < 1$, the MSE diverges, and the time evolution of AMP differs from that of SE.

The instability of AMP stems from the discontinuity of the thresholding function.
As discussed in Sec.~\ref{sec:problem-setting}, $S(x;\lambda R(\cdot))$ becomes discontinuous when $\varepsilon < \sqrt{\lambda}$. In the AMP update, the corresponding condition is $\varepsilon < \sqrt{\chi^{[t]}/\alpha}$, or equivalently, $\chi^{[t]} > \varepsilon^2\alpha$. Defining $\chi_c=\varepsilon^2\alpha$, the AMP update enters the discontinuous regime when $\chi^{[t]} > \chi_c$.
To examine this condition, Fig.~\ref{fig:amp-fixed}(b) shows the time evolution of $\chi^{[t]}/\chi_c$.
While $\chi^{[t]}$ always satisfies $\chi^{[t]} \le \chi_c$ in the setting of $\varepsilon \ge 1$, $\chi^{[t]}$ exceeds $\chi_c$ around $t \approx 4$ in the setting of $\varepsilon < 1$, after which the algorithm heads towards divergence.
These findings suggest that a fixed parameter strategy is insufficient for the log-sum penalty.
Instead, $\varepsilon$ should be set large at the beginning of optimization and should be gradually decreased as $\chi^{[t]}$ shrinks.

\subsection{Adaptive smoothing}

To address algorithmic instability, we propose an adaptive smoothing method that enforces continuity of the thresholding function across iterations.
We treat the smoothing parameter $\varepsilon$ as an iteration-dependent quantity $\varepsilon^{[t]}$.
Then, at each iteration step $t$, we introduce an offset from $\sqrt{\lambda^{[t]}}$ and define $\Delta \varepsilon = \varepsilon^{[t]} - \sqrt{\lambda^{[t]}}$, so that the smoothing parameter at iteration $t$ is written as
\begin{align}
    \varepsilon^{[t]} = \sqrt{\lambda^{[t]}} + \Delta \varepsilon.
    \label{eq:adaptive-smoothing}
\end{align}
When $\Delta \varepsilon \ge 0$, $\varepsilon^{[t]}$ stays on the continuous side.
This setting ensures the thresholding function remains continuous at each iteration step, leading to stable algorithmic behavior.
This update rule requires only a small number of additional scalar operations.
The adaptive smoothing thus introduces only a negligible additional computational cost and does not change the asymptotic computational complexity.

The role of this susceptibility-dependent rule can be understood intuitively by following the progress of the reconstruction.
As indicated in Fig.~\ref{fig:amp-fixed}, the susceptibility $\chi^{[t]}$ decreases as the estimate approaches the zero-error state along a successful reconstruction trajectory.
This decrease can therefore be regarded as a `signature' that the reconstruction is progressing.
The adaptive smoothing uses this signature to retain stronger smoothing when $\chi^{[t]}$ is large early in the iteration and to progressively reduce smoothing as $\chi^{[t]}$ decreases.
In this sense, the adaptive smoothing gradually relaxes the smoothing according to the progress of the reconstruction, while the relation $\varepsilon^{[t]}\ge\sqrt{\chi^{[t]}/\alpha}$ keeps the thresholding function on the continuous side throughout the iteration.

\begin{figure}[tbp]
    \centering
    \includegraphics[width=0.6\columnwidth]{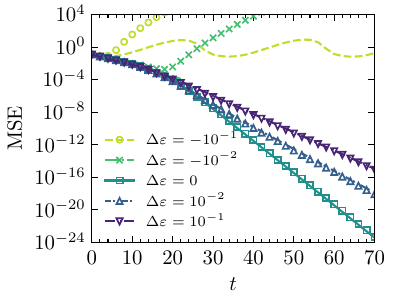}
    \caption{
        Time evolution of MSE with the adaptive smoothing $\varepsilon^{[t]} = \sqrt{\lambda^{[t]}} + \Delta \varepsilon$.
        The symbols and curves represent the results of AMP ($N=10^4$) and SE ($N \to \infty$), respectively.
        Measurement rate and signal density are set to $\alpha=0.5$ and $\rho=0.2$.
    }
    \label{fig:amp-dynamic}
\end{figure}

Numerical simulation confirms that the value of $\Delta \varepsilon$ governs the stability and the convergence speed of the proposed method, as shown in Fig.~\ref{fig:amp-dynamic}.
When $\Delta \varepsilon \ge 0$, MSE decreases monotonically and converges to zero.
In addition, within this range, the smaller $\Delta \varepsilon$ is, the faster the convergence.
This trend is consistent with the intuition that a smaller $\varepsilon$ yields a sharper shrinkage profile near the origin.
Conversely, when $\Delta \varepsilon < 0$, MSE diverges.
Among the stable schedules examined in Fig.~\ref{fig:amp-dynamic}, the boundary schedule corresponding to $\Delta\varepsilon\to0^+$ yields the fastest convergence.

\section{Replica analysis}
\label{sec:replica-analysis}

\subsection{Replica method}

In this section, we investigate the asymptotic behavior of the posterior probability distribution $P_\beta(\bm{x} \mid \bm{y}, \bm{A})$ defined in Eq.~\eqref{eq:posterior-distribution-logsum}.
We study its behavior in the $\beta \to \infty$ limit to evaluate typical reconstruction performance.
We are interested in the asymptotic behavior of the MSE, which is expressed using two macroscopic order parameters, $Q = \|\bm{x}\|_2^2 / N$ and $m = \bm{x}\cdot \bm{x}^0 / N$ as
\begin{align}
    \text{MSE} =
    \mathbb{E}_{\bm{x}, \bm{x}^0}\!\left[
        Q - 2m
    \right] + \rho.
    \label{eq:replica-mse}
\end{align}
To study the system as $N$ increases, we compute the average free energy density $f$ using the replica method.
As $N$ grows, $Q$ and $m$ approach the values determined by the saddle-point of the free energy density
\begin{align}
    f = -\frac{1}{N \beta} \log Z.
\end{align}
Here, $Z$ is the partition function defined by
\begin{align}
    Z = \int d\bm{x} \exp\!\left(
        - \beta R(\bm{x})
        - \frac{1}{2 \sigma^2} \| \bm{A}(\bm{x} - \bm{x}^0) \|_2^2
    \right).
\end{align}
We assume the self-averaging property $f = \mathbb{E}_{\bm A, \bm x^0}[f]$ as $N \to \infty$.
Following the standard replica method, we use the identity $\mathbb{E}_{\bm{A}, \bm{x}^0}[\log Z] = \lim_{n \to 0^+} n^{-1} \log \mathbb{E}_{\bm{A}, \bm{x}^0}[Z^n]$ to express the free energy density as
\begin{align}
    f = - \frac{1}{N \beta} \mathbb{E}_{\bm{A}, \bm{x}^0}[\log Z]
    = - \frac{1}{N \beta} \lim_{n \to 0^+} \frac{1}{n} \log \mathbb{E}_{\bm{A}, \bm{x}^0}[Z^n].
    \label{eq:replica-free-energy-density-to-calculate}
\end{align}

The calculation proceeds as follows.
We first calculate the $n$-th moment $\mathbb{E}_{\bm{A}, \bm{x}^0}[Z^n]$ assuming an integer number of replicas $n \in \mathbb{N}$.
In the thermodynamic limit $N \to \infty$, this moment is evaluated via the saddle-point method with respect to the order parameters.
Assuming we can continue the formula for integer $n$ to positive real values, we then take the limit as $n$ approaches zero.
Subsequently, we take the noiseless limit $\sigma^2 \to 0^+$ to enforce the measurement constraint precisely.
Finally, we take the zero-temperature limit $\beta \to \infty$.

Here we briefly sketch the calculation.
Detailed calculations are provided in Appendix~\ref{sec:apdx:replica-partition-function-result}.
We introduce the following order parameters:
\begin{align}
    Q^{ab} &= \frac{1}{N} \sum_{i=1}^N x_i^a x_i^b, \\
    m^a &= \frac{1}{N} \sum_{i=1}^N x_i^a x_i^0.
\end{align}
Using these variables, $\mathbb{E}_{\bm{A}, \bm{x}^0}[Z^n]$ is simplified as
\begin{align}
    \mathbb{E}_{\bm{A}, \bm{x}^0}[Z^n]
    &=
        \int \prod_{a, b=1}^n dQ^{ab} d\widetilde{Q}^{ab} \int \prod_{a=1}^n dm^a d\widetilde{m}^a \notag
        \\ &\quad
        \exp\!\Bigg(
            \frac{N}{2} \sum_{a,b=1}^n i \widetilde{Q}^{ab} Q^{ab} + N \sum_{a=1}^n i \widetilde{m}^a m^a
            - N n \beta \alpha \Psi - N n \beta \widetilde{\Psi}
        \Bigg),
        \label{eq:replica-partition-function-result}
\end{align}
where
\begin{align}
    n \beta \Psi
    &=
        - \log \mathbb{E}_{y^a, y^0} \exp\!\left(
            -\frac{1}{2 \sigma^2} \sum_{a=1}^n (y^a - y^0)^2
        \right),
        \label{eq:psi-def}
        \\
    n \beta \widetilde{\Psi}
    &=
        - \log \mathbb{E}_{x^0} \int \prod_a dx^a
        \exp\!\Bigg(
            -\frac{1}{2} \sum_{a,b} i\widetilde{Q}^{ab} x^a x^b
            - \sum_a i\widetilde{m}^a x^a x^0
            - \beta \sum_a R(x^a)
        \Bigg).
        \label{eq:tilde-psi-def}
\end{align}
$y^a$ and $y^0$ follow a zero-mean multivariate Gaussian distribution satisfying
\begin{align}
    \mathbb{E}_{y_\mu^a, y_\mu^b} [y_\mu^a y_\mu^b] &= Q^{ab}, \\
    \mathbb{E}_{y_\mu^a, y_\mu^0} [y_\mu^a y_\mu^0] &= m^a, \\
    \mathbb{E}_{y_\mu^0} [(y_\mu^0)^2] &= \rho.
\end{align}
The right-hand side of Eq.~\eqref{eq:replica-partition-function-result} still depends on $n$ being discrete, so we cannot yet analytically continue to $n \to 0^+$.
To proceed, we need to make some assumptions about the order parameters.

\subsection{Replica symmetry solution}

We now assume that the order parameters have the following replica symmetric (RS) structure:~
\begin{align}
    Q^{ab} &= q + \delta_{ab} (Q - q), \label{eq:rs-ansatz-start} \\
    i \widetilde{Q}^{ab} &= \tilde{q} + \delta_{ab} (\widetilde{Q} - \tilde{q}), \\
    m^a &= m, \\
    i \tilde{m}^a &= \tilde{m} \label{eq:rs-ansatz-end}.
\end{align}
Taking the limit $n \to 0^+$ under this assumption yields the following expression of the \textit{RS free energy density}:
\begin{align}
- \beta f_\text{RS}
&=
    \operatorname*{extr}_{Q, q, m, \widetilde{Q}, \tilde{q}, \tilde{m}} \Bigg\{
        \frac{\widetilde{Q}Q}{2} - \frac{\tilde{q}q}{2} + \tilde{m}m
        - \beta \alpha \Psi_\text{RS}
        - \beta \widetilde{\Psi}_\text{RS}
    \Bigg\},
    \label{eq:rs-free-energy}
\end{align}
where
\begin{align}
    \beta \Psi_\text{RS}
    &=
        \frac{1}{2} \left(
            \frac{ q - 2m + \rho }{ \sigma^2 + Q - q }
            + \log\!\left( \sigma^2 + Q - q \right)
            - \log \sigma^2
        \right),
    \label{eq:Psi-RS} \\
    \beta \widetilde{\Psi}_\text{RS}
    &=
        -\mathbb{E}_{x^0} \int D\xi \log \int dx \exp\!\Bigg(
            -\frac{\widetilde{Q}-\tilde{q}}{2}x^2
            + (\sqrt{-\tilde{q}}\,\xi - \tilde{m}x^0)x - \beta R(x)
        \Bigg).
    \label{eq:Psi-tilde-RS}
\end{align}
Here, $\operatorname*{extr}$ means the extremization with respect to the order parameters and their conjugate variables.
The extremum condition of the free energy yields the saddle-point equations
\begin{align}
    -\widetilde{Q} &= - \tilde{q} + \tilde{m}, \\
    -\tilde{q} &= \frac{\alpha (q-2m+\rho)}{(Q-q)^2}, \\
    -\tilde{m} &= \frac{\alpha}{Q-q}, \\
    Q &= \mathbb{E}_{x^0} \int D\xi \left\langle x^2 \right\rangle_{x}, \\
    q &= \mathbb{E}_{x^0} \int D\xi \left\langle x \right\rangle_{x}^2, \\
    m &= \mathbb{E}_{x^0} \int D\xi \left\langle x^0 x \right\rangle_{x}.
\end{align}
Here, $\left\langle \cdot \right\rangle_x$ represents the expectation with respect to $x \sim \pi(x \mid \xi, x^0)$, where
\begin{align}
    \pi(x \mid \xi, x^0)
        &\propto
        \exp\!\left(
            -\frac{\widetilde{Q}-\tilde{q}}{2} x^2 + (\sqrt{-\tilde{q}}\xi - \tilde{m}x^0)x - \beta R(x)
        \right).
\end{align}
The detailed derivations of Eq.~\eqref{eq:rs-free-energy} are presented in Appendix \ref{sec:apdx:rs-free-energy}.

Next, to describe the ground state behavior, we analyze the zero-temperature limit by introducing proper rescalings with respect to $\beta$.
In the $\beta \to \infty$ limit, the order parameters are explained by the scaled parameters $\chi = \beta(Q - q)$, $\widehat{Q} = \beta^{-1}(\widetilde{Q} - \tilde{q})$, $\hat{\chi} = - \beta^{-2} \tilde{q}$, and $\hat{m} = -\beta^{-1} \tilde{m}$.
By applying these rescalings and subsequently taking the $\beta \to \infty$ limit, we obtain the following set of equations:
\begin{align}
    \widehat{Q} &= \hat{m} = \frac{\alpha}{\chi}, \label{eq:SPE-hat-Q} \\
    \hat{\chi} &= \frac{\alpha(Q-2m+\rho)}{\chi^2}, \label{eq:SPE-hat-chi} \\
    Q &= \mathbb{E}_{x^0} \int D\xi\, (x^*)^2, \label{eq:SPE-Q} \\
    \chi &= \mathbb{E}_{x^0} \int D\xi\, \frac{\xi x^*}{\sqrt{\hat{\chi}}}, \\
    m &= \mathbb{E}_{x^0} \int D\xi\, x^0 x^*, \label{eq:SPE-m}
\end{align}
where
\begin{align}
    x^*
        &= S\!\left( h_\text{RS}; \frac{1}{\widehat{Q}} R(\cdot) \right),
        \label{eq:RS-effective-estimator} \\
    h_\text{RS}
        &= \frac{\hat{m}}{\widehat{Q}} x^0 + \frac{\sqrt{\hat{\chi}}}{\widehat{Q}} \xi
        = x^0 + \sqrt{\frac{Q - 2m + \rho}{\alpha}} \xi
        .
        \label{eq:RS-effective-field}
\end{align}
When $\varepsilon \geq \widehat{Q}^{-1/2}$, this set of equations coincides with the fixed-point conditions of the SE recursion in Eqs.~\eqref{eq:SE-effective-field}--\eqref{eq:SE-chi}.
The correspondence is discussed in detail in Appendix~\ref{sec:apdx:rs-se-correspondence}.
We emphasize that this derivation establishes a correspondence between stationary conditions only.
The equations above do not specify an iteration rule and therefore do not reproduce the transient trajectory of SE.

\subsection{Stability of perfect reconstruction}

We investigate the stability of the fixed-point solution corresponding to exact recovery based on the zero-temperature equations given in Eqs.~\eqref{eq:SPE-hat-Q}--\eqref{eq:SPE-m}.
In the exact-recovery limit ($\mathrm{MSE} \to 0^+$), the order parameters converge to $Q \to \rho$, $m \to \rho$, $\chi \to 0^+$, and $\widehat{Q} = \hat{m} \to \infty$.
Consequently, only $\hat{\chi}$ converges to a nontrivial stable fixed point $\hat{\chi}^\ast = O(1)$.
Assuming the convexity condition
\begin{align}
    \varepsilon > \sqrt{\frac{\chi}{\alpha}},
    \label{eq:continuity-log-sum-RS}
\end{align}
the self-consistent equation that determines $\hat{\chi}^*$ is given as
\begin{align}
    \varepsilon^2 \hat{\chi} &=
        \frac{1}{\alpha} \Bigg(
            2 (1 - \rho) \left(
                \left( \varepsilon^2 \hat{\chi} + 1 \right) H \!\left( \frac{1}{ \sqrt{\varepsilon^2 \hat{\chi}} }
            \right)
            - \sqrt{\varepsilon^2 \hat{\chi}}
            \phi\!\left(\frac{1}{ \sqrt{\varepsilon^2 \hat{\chi}} }\right)
        \right)
        \notag \\ & \qquad
        + \rho \left(
            \varepsilon^2 \hat{\chi} + \varepsilon^2 \int D \xi \frac{1}{ \left( |\xi| + \varepsilon \right)^2 }
        \right)
        \Bigg).
    \label{eq:self-consistent-hat-chi}
\end{align}
The detailed derivation is provided in Appendix~\ref{sec:apdx:linear-stability}.

By examining the local stability of the fixed point $\hat{\chi}^*$, we determine the linear stability boundary $\alpha_c(\rho)$ as
\begin{align}
    \alpha_c(\rho) = \rho + 2(1 - \rho) H\!\left( \frac{1}{ \sqrt{\varepsilon^2 \hat{\chi}^*} }\right). \label{eq:alpha_c-log-sum}
\end{align}
Here, $H(x) = \int_x^\infty D \xi$ is the complementary cumulative distribution function (CCDF) of the standard normal distribution.
$\alpha > \alpha_c(\rho)$ is necessary for the exact-recovery state to be stable; in the region $\alpha < \alpha_c(\rho)$, exact recovery ($\mathrm{MSE}=0$) is unstable and is therefore unreachable.
The resulting phase boundary on the $\alpha$--$\rho$ plane is illustrated in Sec.~\ref{sec:phase-diagram}.

Moreover, Eqs.~\eqref{eq:continuity-log-sum-RS}--\eqref{eq:alpha_c-log-sum} suggest that the log-sum penalty allows for exact recovery up to the IT threshold, $\alpha_c = \rho$.
Considering the exact-recovery limit, where $\widehat{Q}_\text{RS} \to \infty$ and $\chi_\text{RS} \to 0^+$, the right-hand side of Eq.~\eqref{eq:continuity-log-sum-RS} converges to $0^+$.
As a result, this convergence allows $\varepsilon$ to approach an arbitrarily small positive value without violating the convexity condition.
In addition, as $\varepsilon \to 0^+$, the second term in Eq.~\eqref{eq:alpha_c-log-sum} vanishes, leading to the limit $\alpha_c(\rho) \to \rho^+$.
These observations imply that, in principle, $\text{MSE}=0$ is reachable for all $\alpha > \rho$ when using the log-sum penalty.
However, the actual success region is restricted by the emergence of a finite-MSE fixed point, as discussed in Sec.~\ref{sec:phase-diagram}.

\subsection{Stability of RS solution}

We examine the validity of the replica symmetric (RS) ansatz, which assumes that all replicas are treated identically. To do this, we derive the de Almeida–Thouless (dAT) stability condition.
This condition refers to the case in which the RS fixed point described above remains stable under perturbations that violate the symmetry between replicas. For the model discussed in this paper, the dAT stability condition is
\begin{align}
    \alpha > \mathbb{E}_{x^0} \int D\xi\, S'\!\left( x^0 + \frac{\sqrt{\hat{\chi}_\text{RS}}}{\widehat{Q}_\text{RS}} \xi; \frac{1}{\widehat{Q}_\text{RS}} R(\cdot) \right)^{2}.
    \label{eq:dat-condition}
\end{align}
Here, $\widehat{Q}_\text{RS}$ and $\hat{\chi}_\text{RS}$ are the RS fixed points of $\widehat{Q}$ and $\hat{\chi}$, respectively.
The detailed derivation is provided in Appendix \ref{sec:apdx:1rsb-free-energy}.
In the $\text{MSE} \to 0^+$ limit, the right-hand side of Eq.~\eqref{eq:dat-condition} coincides with $\alpha_c(\rho)$.
Consequently, we see that in the region $\alpha > \alpha_c(\rho)$, the RS fixed point remains stable.

\section{Phase diagram and numerical validations}
\label{sec:phase-diagram}

In this section, we visualize the local-stability thresholds for fixed $\varepsilon$ and the algorithmic threshold for adaptive smoothing and compare them with the reconstruction threshold for $\ell_1$ minimization, BO threshold, and IT threshold.

\begin{figure}[tbp]
    \centering
    \includegraphics[width=0.5\columnwidth]{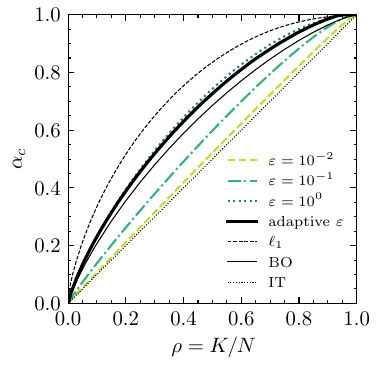}
    \caption{
        Typical reconstruction thresholds $\alpha_c$ as a function of the signal density $\rho$.
        The colored lines represent the local-stability thresholds of the log-sum penalty for fixed parameter values $\varepsilon = 10^{-2}, 10^{-1}$, and $10^0$, while the thick solid black line corresponds to the algorithmic threshold of adaptive smoothing.
        For comparison, the reconstruction threshold for $\ell_1$ minimization ($\ell_1$), the BO threshold (BO), and the IT threshold (IT) are also plotted.
    }
    \label{fig:phase-diagram}
\end{figure}

In Fig.~\ref{fig:phase-diagram}, we plot the local-stability threshold $\alpha_c$, derived in Sec.~\ref{sec:replica-analysis}, as a function of $\rho$, along with the algorithmic threshold obtained using adaptive smoothing.
The boundary for the adaptive smoothing is determined numerically as follows.
Specifically, for each $\rho$, we use binary search for the minimum $\alpha$ for which $\text{MSE} < 10^{-3}$ is achieved.
For each $\rho$ and $\alpha$, we iterate the SE equations given in Eqs.~\eqref{eq:SE-effective-field}--\eqref{eq:SE-chi} until either the iteration count reaches $t_\text{max} = 10^{3}$,
the MSE falls below $10^{-4}$,
or the MSE exceeds $10^{4}$.
The SE dynamics are initialized with
$\text{MSE}^{[0]}=\rho$ and $\chi^{[0]}=1$,
which corresponds to
$\hat{\bm{x}}^{[0]} = \bm{0}$ and $\chi^{[0]}=1$ for the AMP algorithm,
to verify whether $\text{MSE}=0$ is typically achievable from such an uninformed initial condition.
We hereafter refer to this threshold as the \textit{adaptive-smoothing threshold}.
For comparison, we display the reconstruction threshold for $\ell_1$ minimization, the BO threshold, and the IT threshold.

The phase diagram reveals two key points that clarify the observed behavior.
First, and as discussed in Sec.~\ref{sec:replica-analysis}, the local-stability threshold converges to $\alpha_c=\rho$, which coincides with the IT threshold in the present setting, as $\varepsilon\to0^+$.
Second, the adaptive-smoothing threshold lies slightly above the BO threshold.
These two observations reveal the separation between the local-stability threshold for fixed $\varepsilon$ and the adaptive-smoothing threshold.
The phase in this gap---the region between the local-stability threshold and adaptive-smoothing threshold---is characterized by the emergence of a finite-MSE fixed point, as we discuss below.
In the rest of the paper, we refer to this region as the \textit{hard phase}.

\begin{figure*}[tbp]
    \centering
    \includegraphics[width=1\columnwidth]{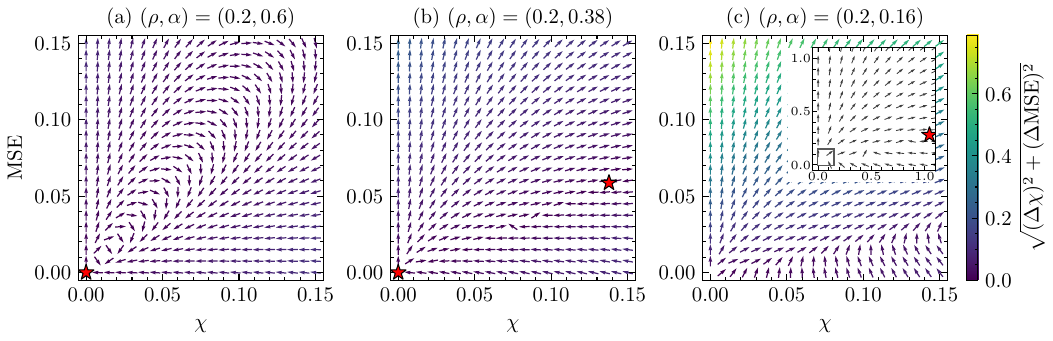}
    \caption{
        Vector field of the SE dynamics with adaptive smoothing on the $\text{MSE}$--$\chi$ plane.
        (a) Easy phase, $(\rho, \alpha) = (0.2, 0.6)$.
        (b) Hard phase, $(\rho, \alpha) = (0.2, 0.38)$.
        (c) Impossible phase, $(\rho, \alpha) = (0.2, 0.16)$.
        Red stars denote locally stable fixed points.
        The color scale shows the velocity magnitude $\sqrt{(\mathrm{MSE}^{[t+1]}-\mathrm{MSE}^{[t]})^2+(\chi^{[t+1]}-\chi^{[t]})^2}$.
    }
    \label{fig:se-field}
\end{figure*}

We next investigate the mechanism underlying the failure of AMP with adaptive smoothing to achieve the exact recovery.
To understand dynamics across phases, we visualize the SE dynamics from Eqs.~\eqref{eq:SE-effective-field}--\eqref{eq:SE-chi}.
We show them as a vector field on the $\text{MSE}$--$\chi$ plane in Fig.~\ref{fig:se-field}.
In the easy phase (Fig.~\ref{fig:se-field}(a)), corresponding to the region above the adaptive-smoothing threshold, the basin of attraction of the zero-MSE fixed point extends over a wide region.
It directs trajectories toward a fixed point where $\text{MSE} = \chi = 0$.
In the impossible phase (Fig.~\ref{fig:se-field}(c)), i.e., below the IT threshold, the zero-MSE fixed point is no longer stable, while a locally stable fixed point with $\mathrm{MSE}>0$ is present.
This finite-MSE fixed point attracts SE trajectories within its basin of attraction, thereby preventing them from reaching exact recovery.
Between these, the hard phase (Fig.~\ref{fig:se-field}(b)) is characterized by the coexistence of this finite-MSE fixed point and the locally stable zero-MSE fixed point.
The fixed point at the origin remains locally stable, and the SE flow still contains trajectories directed toward $\mathrm{MSE}=\chi=0$.
This zero-MSE fixed point, however, can be reached only when the initial state belongs to the basin of attraction of it.
Initial states outside this basin are instead attracted to the finite-MSE fixed point.
In particular, for the uninformed initial conditions considered here, the SE flow is attracted to the finite-MSE fixed point, and the AMP dynamics typically exhibit the corresponding behavior.
Thus, although the exact-recovery state remains locally stable in the hard phase, the algorithm generally fails to reach it from the initialization used in this study.

The persistence of the hard phase under adaptive smoothing can be understood by considering how the finite-MSE state interacts with the adaptive smoothing rule.
According to Eq.~\eqref{eq:SE-effective-field}, finite MSE induces a Gaussian fluctuation with standard deviation $\sqrt{\mathrm{MSE} / \alpha}$ in the effective field.
As $\alpha$ decreases, finite MSE produces a larger fluctuation and thereby possibly increases the MSE after thresholding through Eq.~\eqref{eq:SE-mse}.
This feedback allows a finite-MSE state to remain self-consistent in the hard phase.
The adaptive smoothing rule is coupled to the SE dynamics through the current value of $\chi$.
When the SE dynamics converge to the zero-MSE fixed point, $\varepsilon$ gradually decreases as $\chi$ shrinks toward $0$.
However, when the SE dynamics converge to a finite-MSE fixed point, the parameter converges to a finite value $\varepsilon^* = \sqrt{\chi^* / \alpha} > 0$.
Therefore, once the SE dynamics converge to the finite-MSE fixed point, the adaptive update of $\varepsilon$ also becomes stationary.
Thus, adaptive smoothing alone does not provide a mechanism for driving a trajectory toward the finite-MSE fixed point out of its basin of attraction.

\begin{figure}[tbp]
    \centering
    \includegraphics[width=0.5\columnwidth]{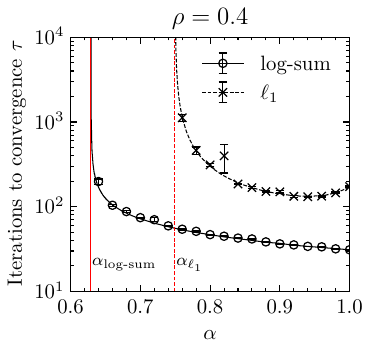}
    \caption{
        Number of iterations for AMP to converge as a function of measurement rate $\alpha$ for $\rho=0.4$.
        Symbols represent the averages over $10$ independent results with $N=10^{4}$, and error bars indicate the standard error.
        Curves correspond to the SE predictions in the $N\to\infty$ limit.
    }
    \label{fig:convergence-time}
\end{figure}

To further confirm the practical benefits of the proposed adaptive smoothing, we examine how many iterations the AMP algorithm needs to converge.
This performance metric is especially important in real-world scenarios, since excessively slow convergence limits the practical applicability of an algorithm, even when it ultimately reconstructs the true signal $\bm{x}^0$.
Figure~\ref{fig:convergence-time} displays how many iterations the AMP algorithm needs to converge for $\rho = 0.4$ and $\alpha$ between $0.6$ and $1$.
We generate $10$ independent instances of Gaussian measurement matrices $\bm{A}$ and signals $\bm{x}^0$ of dimension $N = 10^4$, drawn from the Bernoulli--Gaussian distribution $P_{\mathrm{true}}(\bm{x}^0)$.
The typical number of nonzero components is $K = 4 \times 10^3$.
For each instance, the AMP algorithm is initialized at $\hat{\bm{x}}^{[0]} = \bm{0}$ and terminated when $\text{MSE} < 10^{-10}$ is satisfied.
We define the convergence time $\tau$ as the number of iterations required to meet this criterion.
If convergence does not occur within $t_\text{max} = 5 \times 10^4$ iterations, $\tau$ is set to $\infty$.
For $\alpha < 0.6$, none of the tested instances converged to the true signal within $t_{\max}$.
In the range where the algorithm converges, the log-sum version requires approximately one order of magnitude fewer iterations than the $\ell_1$ version.
Over most of that range, $\tau$ stays nearly constant.
It increases only slightly as $\alpha$ approaches the reconstruction threshold $\alpha_c = \alpha_\text{log-sum}$.

\section{Discussion and Conclusion}
\label{sec:discussion}

Our results identify two asymptotic performance limits for log-sum minimization.
The first is the adaptive-smoothing threshold, which characterizes the algorithmic threshold of the proposed method.
This threshold lies below the $\ell_1$ threshold and slightly above the BO threshold.
The second is the local-stability threshold for a fixed $\varepsilon$ in the limit $\varepsilon \to 0^+$, which coincides with the IT threshold $\alpha_c = \rho$.
The gap between the adaptive-smoothing threshold and the IT threshold defines the hard phase, where the exact-recovery solution remains locally stable but is not typically reached by the proposed algorithm from an uninformed initial condition.

Similar gaps between the local-stability threshold and the algorithmic threshold have been reported for other algorithms for nonconvex sparse reconstruction, including AMP with SCAD and MCP~\cite{Sakata2021-du}, its extension to approximate survey propagation (ASP)~\cite{Gu2025-wr}, and ASP for $\ell_0$ regularization~\cite{Barbier2025-ob}.
A closely related behavior has also been observed for compressed sensing based on Bayesian linear regression with a horseshoe prior~\cite{Nagano2023-kq,Nagano2024-ig}, where decreasing the global shrinkage parameter (denoted by $\tau$ in \cite{Nagano2024-ig}) enlarges the exact-recovery region but produces a \textit{hard phase}.
This behavior closely parallels the role of the smoothing parameter $\varepsilon$ in the log-sum penalty.
The connection is also consistent with the heavy tail of the horseshoe prior that decays as a power law \cite{Carvalho2010-rz}, although the horseshoe prior and the log-sum penalty differ near the origin; a more direct Bayesian counterpart of the log-sum penalty is provided by the generalized double Pareto distribution~\cite{Armagan2013-js}.
These results suggest that a similar separation between the local-stability threshold and the algorithmic threshold may commonly arise for nonconvex sparse reconstruction methods, while a quantitative comparison with other nonconvex penalties is left for future work.

For the present log-sum problem, the fixed-point structure underlying the hard phase suggests two possible routes for improving the algorithmic threshold.
One is to introduce a more informative initialization.
As shown in Fig.~\ref{fig:se-field}, the perfect-reconstruction fixed point remains locally stable in this region, whereas the uninformed initial condition used in this study is attracted to a competing finite-MSE fixed point.
This suggests that an informative initialization could potentially reduce the hard phase by placing the initial state within the basin of attraction of the perfect-reconstruction fixed point.
However, such an initialization requires an informative estimate of the unknown signal to be available in advance, which may itself require additional prior information or a separate reconstruction procedure.
Alternative continuation strategies may provide another possibility.
The present adaptive-smoothing algorithm adjusts $\varepsilon$ to maintain continuity of the thresholding function, but it does not explicitly control which locally stable fixed point the SE dynamics reach.
A continuation strategy that also steers the trajectory toward the basin of the perfect-reconstruction fixed point may therefore improve the algorithmic threshold.
One possible approach along this direction is the nonconvexity control (NCC) proposed in Refs.~\cite{Sakata2018-ey, Sakata2021-du}.
NCC updates the nonconvexity parameter after the AMP dynamics become sufficiently stationary so as to guide the trajectory toward the zero-MSE fixed point, whereas our adaptive smoothing updates the smoothing parameter at every iteration according to the current susceptibility to maintain continuity of the thresholding function.
A systematic comparison between these control strategies is left for future work.

From a broader perspective, the NCC and the present adaptive smoothing can both be placed within the wider class of continuation methods.
Continuation methods for nonconvex optimization can be broadly classified into two strategies.
One gradually modifies the optimization problem according to a predetermined parameter schedule, and the other adapts the continuation parameter according to the current estimation or optimization state.
In the former strategy, a parameter controlling the degree of nonconvexity is externally varied, thereby moving gradually from a smoother or more tractable problem toward the target problem with stronger nonconvexity.
This strategy is used, for example, in the smoothed $\ell_0$ method \cite{Mohimani2009-lg} and successive concave sparsity approximation \cite{Malek-Mohammadi2016-ix}.
In the latter strategy, the timing or magnitude of the continuation update is determined using the current estimate, the optimization error, or state variables computed within the algorithm.
Examples of such a strategy include the adaptive update of the parameter in IRLS \cite{Daubechies2010-sj} and the adjustment of the parameter based on the duality gap \cite{Hadj-Selem2018-zk}.
The adaptive smoothing proposed in this study belongs to the latter class of state-dependent continuation, in which the continuation parameter is updated at every AMP iteration so as to track the susceptibility-dependent continuity boundary of the scalar thresholding problem.

There are several limitations to the results presented in this paper.
First, the phase boundaries reported in this study apply to noiseless linear observations with i.i.d.~Gaussian measurement matrices and componentwise i.i.d.~Bernoulli--Gaussian signals in the large-system limit, $N,M\to\infty$ at fixed $M/N=\alpha$.
The noiseless assumption is essential to our use of exact recovery ($\mathrm{MSE}=0$) as the success criterion.
The effects of measurement noise also need to be examined.
This would require analyzing noise sensitivity and how reconstruction performance depends on the signal-to-noise ratio \cite{Wu2012-oe}.
Second, the SE predictions and the numerical phase boundaries derived here cannot be transferred unchanged to correlated or nonzero-mean measurement matrices.
Structured signals, for example, would require an analysis under different assumptions.

These limitations lead to a major future challenge:~relaxing assumptions regarding measurement matrices, signal structures, and observation models.
In many real-world problems, measurement matrices may have nonzero means or strong correlations between columns or rows.
In such conditions, standard AMP may lose its convergence properties, and its SE may no longer accurately describe the algorithm’s behavior.
Enhanced approaches such as orthogonal AMP (OAMP) \cite{Ma2017-tk} and vector AMP (VAMP) \cite{Rangan2017-vh}
provide a natural direction for extending the current framework to a broader class of measurement matrices, including unitarily invariant and right-orthogonally invariant matrices.
Such an extension would allow us to examine whether adaptive smoothing still achieves a wider reconstruction region than $\ell_1$-norm minimization even beyond the ideal i.i.d.\ Gaussian setting, and whether the separation between local stability and algorithmic thresholds is maintained.
A key challenge is whether the adaptive smoothing strategy can be incorporated into the estimation steps of these algorithms.
In the estimation steps of the standard AMP framework used in this study, maintaining the continuity of the log-sum threshold function can be achieved by simply adjusting the sensitivity $\chi$.
In OAMP and VAMP, however, the estimation steps are characterized by different state quantities.
Extending adaptive smoothing to these algorithms would therefore require reformulating the continuity condition of the log-sum thresholding function in terms of the corresponding algorithm-specific state quantities.

Addressing various types of signal and measurement models is also indispensable to enhance practical applicability.
In image processing and communications, block sparsity or tree-structured sparsity is often more appropriate than independent sparsity.
Therefore, combining the log-sum penalty with grouped regularization or total-variation (TV) regularization and analyzing the changes in metastability under structured priors would constitute an important future direction.
Since the scalar thresholding step examined in this study is replaced by a \textit{coupled} estimation step under such settings, it is necessary to identify the conditions under which the resulting coupled estimation step loses algorithmic stability.
This would also require reformulating the adaptive-smoothing rule by identifying an appropriate stability condition for the coupled estimator and controlling the smoothing parameter accordingly.
For noisy measurements, retaining finite $\sigma^2$ in the posterior formulation provides a natural starting point.
A particularly relevant question is how the hard phase and competing finite-error fixed points evolve when exact reconstruction is replaced by noisy reconstruction, and whether adaptive smoothing can still guide the dynamics toward lower-error solutions.

In addition to the log-sum penalty focused on in this study, extending the scope of performance evaluation to other, more modern penalty functions, such as the one proposed in \cite{Furuhashi2026WEEP}, is also an interesting topic for future research.
In particular, analyzing the possibility of an emerging hard phase, the conditions where true sparse signals can be recovered in compressed sensing, and how closely their typical performance can approach the BO threshold would be an important contribution to the development of sparse modeling.

\section*{Acknowledgments}

K.M.~thanks M.~Doi for valuable discussions.
We received financial support from programs for bridging the gap between R\&D and IDeal society (Society 5.0) and Generating Economic and social value (BRIDGE) and Cross-ministerial Strategic Innovation Promotion Program (SIP) from the Cabinet Office (No.~23836436).
The work of F.R.T.\ was supported by the “National Centre for HPC, Big Data and Quantum Computing”, Project CN\_00000013, CUP B83C22002940006, NRRP Mission 4 Component 2 Investment 1.4, Funded by the European Union - NextGenerationEU.

\appendix
\section[Derivation of Replicated Partition Function]%
{Derivation of Eq.~\eqref{eq:replica-partition-function-result}}%
\label{sec:apdx:replica-partition-function-result}

This appendix provides a detailed derivation of Eq.~\eqref{eq:replica-partition-function-result}. We start from the following replicated partition function:
\begin{align}
    &
        \mathbb{E}_{\bm{A}, \bm{x}^0}[Z^n]
        \notag \\
    &=
        \mathbb{E}_{\bm{A}, \bm{x}^0} \int \prod_{a=1}^n d\bm{x}^a \exp\!\Bigg(
            -\beta \sum_{a=1}^n R(\bm{x}^a)
            - \frac{1}{2 \sigma^2} \sum_{a=1}^n \| \bm{A}(\bm{x}^a - \bm{x}^0) \|_2^2
        \Bigg)
    \notag \\
    &=
        \mathbb{E}_{\bm{x}^0} \int \prod_{a=1}^n d\bm{x}^a
        \exp\!\Bigg(
            -\beta \sum_{a=1}^n R(\bm{x}^a)
        \Bigg)
        \underline{
            \mathbb{E}_{\bm{A}}
            \exp\!\Bigg(
                - \frac{1}{2 \sigma^2} \sum_{a=1}^n \| \bm{A}(\bm{x}^a - \bm{x}^0) \|_2^2
            \Bigg)
        }_{(P_1)}
    .
    \label{eq:replica-partition-function-start}
\end{align}

We first introduce the following auxiliary variables:
\begin{align}
    y_\mu^a = \sum_{i=1}^N A_{\mu i} x_i^a,
    \qquad
    y_\mu^0 = \sum_{i=1}^N A_{\mu i} x_i^0
    \qquad (\text{for} ~ \mu=1,\dots,M, ~ a=1,\dots,n).
\end{align}
The $\bm{A}$-dependent part $(P_1)$ can be rewritten as
\begin{align}
    \mathbb{E}_{\bm{A}}\exp\!\left(
        -\frac{1}{2 \sigma^2}\sum_{a=1}^n \| \bm{A}(\bm{x}^a-\bm{x}^0)\|_2^2
    \right)
    &=
    \prod_{\mu=1}^M
    \mathbb{E}_{y_\mu^a,y_\mu^0 \sim P_y}
    \exp\!\left(
        -\frac{1}{2 \sigma^2}\sum_{a=1}^n (y_\mu^a-y_\mu^0)^2
    \right) \notag \\
    &=
    \exp\!\left(
        - M n \beta \Psi
    \right),
\end{align}
where $\Psi$ is defined in Eq.~\eqref{eq:psi-def}, and $P_y$ is a multivariate Gaussian distribution with zero mean and second moments given by
\begin{align}
    \mathbb{E}_{\{y_\mu^a, y_\mu^0\} \sim P_y}\!\left[ y_\mu^a y_\mu^b \right]
    &= \frac{1}{N}\sum_{i=1}^N x_i^a x_i^b
    \eqqcolon Q^{ab},
    \label{eq:apdx:cov-yab}
    \\
    \mathbb{E}_{\{y_\mu^a, y_\mu^0\} \sim P_y}\!\left[ y_\mu^a y_\mu^0 \right]
    &= \frac{1}{N}\sum_{i=1}^N x_i^a x_i^0
    \eqqcolon m^a,
    \label{eq:apdx:cov-ya0}
    \\
    \mathbb{E}_{\{y_\mu^a, y_\mu^0\} \sim P_y}\!\left[ (y_\mu^0)^2 \right]
    &= \frac{1}{N}\sum_{i=1}^N (x_i^0)^2
    \eqqcolon \rho'.
    \label{eq:apdx:cov-y00}
\end{align}
In the large system size limit $N \to \infty$, the value of $\rho'$ becomes self-averaging, and the following identity holds:
\begin{align}
    \rho' = \mathbb{E}_{x^0}[(x^0)^2] = \rho.
\end{align}
Thus, we replace it with $\rho$ and treat it as a constant hereafter.
Ensuring that other moments take defined values is achieved by redefining the order parameters as random variables following the distributions below:
\begin{align}
    Q^{ab}
    &\sim \delta\!\left(
        NQ^{ab}-\sum_{i=1}^N x_i^a x_i^b
    \right),
    \\
    m^a
    &\sim
    \delta\!\left(
        Nm^a-\sum_{i=1}^N x_i^a x_i^0
    \right).
\end{align}
The resultant replicated partition function is as follows.
\begin{align}
    \mathbb{E}_{\bm{A}, \bm{x}^0}[Z^n]
    &=
        \int \prod_{a, b} d Q^{a b}
        \int \prod_{a} d m^{a}
        \exp\!\left(
            - M n \beta \Psi
        \right)
        \underline{
            \mathbb{E}_{\bm{x}^0}
            \int \prod_{a=1}^n d\bm{x}^a
            \exp\!\left(
                -\beta \sum_{a=1}^n R(\bm{x}^a)
            \right)
        }
        \notag \\ & \quad
        \underline{
            \prod_{a,b}\delta\!\left(
                NQ^{ab}-\sum_{i=1}^N x_i^a x_i^b
            \right)
            \prod_a\delta\!\left(
                Nm^a-\sum_{i=1}^N x_i^a x_i^0
            \right)
        }_{(P_2)}.
\end{align}

Next, we use Fourier integral representations of the delta functions:
\begin{align}
    \delta\!\left(
        NQ^{ab}-\sum_{i=1}^N x_i^a x_i^b
    \right)
    &=
    \int d\widetilde{Q}^{ab}
    \exp\!\left(
        \frac{i\widetilde{Q}^{ab}}{2}
        \left(
            NQ^{ab}-\sum_{i=1}^N x_i^a x_i^b
        \right)
    \right),
    \\
    \delta\!\left(
        Nm^a-\sum_{i=1}^N x_i^a x_i^0
    \right)
    &=
    \int d\tilde{m}^{a}
    \exp\!\left(
        i\tilde{m}^{a}
        \left(
            Nm^a-\sum_{i=1}^N x_i^a x_i^0
        \right)
    \right).
\end{align}
Substituting these expressions and collecting all terms depending on $x_i^a$ and $x_i^0$, the $(P_2)$ part is calculated as
\begin{align}
    &
    \mathbb{E}_{\bm{x}^0}
    \int \prod_{a=1}^n d\bm{x}^a
    \exp\!\left(
        -\beta \sum_{a=1}^n R(\bm{x}^a)
    \right)
    \prod_{a,b}\delta\!\left(
        NQ^{ab}-\sum_{i=1}^N x_i^a x_i^b
    \right)
    \prod_a\delta\!\left(
        Nm^a-\sum_{i=1}^N x_i^a x_i^0
    \right)
    \notag\\
    &=
    \int \prod_{a,b} d\widetilde{Q}^{ab}\int \prod_a d\tilde{m}^a
    \exp\!\left(
        \frac{N}{2}\sum_{a,b} i\widetilde{Q}^{ab} Q^{ab}
        +N\sum_a i\tilde{m}^a m^a
    \right)
    \notag\\
    &\qquad\times
    \prod_{i=1}^N
    \mathbb{E}_{x_i^0}\int \prod_{a=1}^n dx_i^a
    \exp\!\left(
        -\frac{1}{2}\sum_{a,b} i\widetilde{Q}^{ab} x_i^a x_i^b
        -\sum_a i\tilde{m}^a x_i^a x_i^0
        -\beta \sum_{a=1}^n R(x_i^a)
    \right).
\end{align}
Because the single-site integral is identical for all $i \in \{1, \dots, N\}$, the $\{x^a\}$-dependent part can be rewritten as the following single-site contribution:
\begin{align}
    \prod_{i=1}^N
    \mathbb{E}_{x_i^0}\int \prod_{a=1}^n dx_i^a
    \exp\!\left(
        -\frac{1}{2}\sum_{a,b} i\widetilde{Q}^{ab} x_i^a x_i^b
        -\sum_a i\tilde{m}^a x_i^a x_i^0
        -\beta \sum_{a=1}^n R(x_i^a)
    \right)
    &=
    \exp\!\left(
        - N n \beta \widetilde{\Psi}
    \right),
\end{align}
with $\widetilde{\Psi}$ defined in Eq.~\eqref{eq:tilde-psi-def}.
Combining summation terms and the single-site contribution yields Eq.~\eqref{eq:replica-partition-function-result}.

\section[Derivation of RS free energy density]%
{Derivation of Eq.~\eqref{eq:rs-free-energy}}%
\label{sec:apdx:rs-free-energy}

This appendix gives the derivation of Eq.~\eqref{eq:rs-free-energy} from Eq.~\eqref{eq:replica-partition-function-result} under the RS ansatz given in Eqs.~\eqref{eq:rs-ansatz-start}--\eqref{eq:rs-ansatz-end}.

The $\Psi$ term (Eq.~\eqref{eq:psi-def}) is simplified as follows.
The random variables $y^a - y^0$ can be expressed as the linear combination of $n+1$ standard Gaussian variables $\xi_0, \{\xi_1^{a}\}_{a=1, \dots, n}$:
\begin{align}
    y^a - y^0 = \sqrt{q - 2m + \rho} \, \xi_0 + \sqrt{Q - q} \, \xi_1^{a}.
\end{align}
This decomposition yields
\begin{align}
    n \beta \Psi = - \log
        \int D \xi_0
        \left(
            \int D \xi_1
            \exp\!\left(
                -\frac{1}{2 \sigma^2} \left(
                    \sqrt{q - 2m + \rho} \, \xi_0 + \sqrt{Q - q} \, \xi_1
                \right)^2
            \right)
        \right)^n
        .
\end{align}
Here, we used $\prod_a\int D \xi_1^a f(\xi_1^a) = \left(\int D \xi_1 f(\xi_1)\right)^n$.
Then, by repeatedly applying the following Gaussian integral formula,
\begin{align}
    \int D\xi
    \exp \!\left(-\frac{\lambda}{2} (A \xi + B)^2 \right)
    =
    \left(1 + \lambda A^2\right)^{- 1/2} \exp \!\left( -\frac{1}{2} \frac{\lambda B^2}{1 + \lambda A^2} \right),
\end{align}
we obtain the following expression for small $n$:
\begin{align}
    n \beta \Psi
    &=
        - \log
            \!\left( 1 + n \frac{q - 2m + \rho}{\sigma^2 + Q - q} \right)^{-\frac{1}{2}}
            \left(1 + \frac{Q - q}{\sigma^2}\right)^{-\frac{n}{2}}
        \notag \\
    &=
        \frac{n}{2} \left(
            \frac{q - 2m + \rho}{\sigma^2 + Q - q}
            + \log\!\left(1 + \frac{Q - q}{\sigma^2}\right)
        \right)
        + O(n^2).
\end{align}

Next, the $\widetilde{\Psi}$ term (Eq.~\eqref{eq:tilde-psi-def}) is simplified as follows.
By applying the RS ansatz, the summation terms become
\begin{align}
    \sum_{a,b} i\widetilde{Q}^{ab} x^a x^b
        &= \tilde{q} \left( \sum_a x^a \right)^2 + (\widetilde{Q} - \tilde{q}) \sum_a (x^a)^2, \\
    \sum_a i\tilde{m}^a x^a x^0
        &= \tilde{m} x^0 \sum_a x^a,
\end{align}
which gives
\begin{align}
    & n \beta \widetilde{\Psi}
    \notag \\
    &= -\log \mathbb{E}_{x^0} \int \prod_a dx^a
    \notag \\ & \qquad
    \exp \Biggl(
        -\frac{\widetilde{Q}-\tilde{q}}{2} \sum_a (x^a)^2
        - \frac{\tilde{q}}{2} \left( \sum_a x^a \right)^2
        - \tilde{m} x^0 \sum_a x^a
        - \beta \sum_a R(x^a)
    \Biggr) \notag \\
    &= -\log \mathbb{E}_{x^0} \int D \xi \int \prod_a dx^a
    \prod_a \exp \!\left(
        -\frac{\widetilde{Q}-\tilde{q}}{2} (x^a)^2 + (\sqrt{-\tilde{q}} \xi - \tilde{m} x^0) x^a - \beta R(x^a)
    \right).
\end{align}
In the last line, we apply the Hubbard--Stratonovich transformation
\begin{align}
    \exp \!\left( -\frac{\tilde{q}}{2} \left( \sum_a x^a \right)^2 \right) = \int D\xi \prod_a \exp \!\left( \sqrt{-\tilde{q}} x^a \xi \right).
\end{align}
Now that the $n$-fold integral with respect to $x^a$ is identical to the $n$-th power of the same integral, we obtain
\begin{align}
    n \beta \widetilde{\Psi}
    &=
        -\log \mathbb{E}_{x^0} \int D\xi \left(
            \int dx \exp \!\left(
                -\frac{\widetilde{Q}-\tilde{q}}{2} x^2 + (\sqrt{-\tilde{q}} \xi - \tilde{m} x^0) x - \beta R(x)
            \right)
        \right)^n \notag \\
    &=
        -n \mathbb{E}_{x^0} \int D\xi \log \int dx \exp \!\left(
            -\frac{\widetilde{Q}-\tilde{q}}{2} x^2 + (\sqrt{-\tilde{q}} \xi - \tilde{m} x^0) x - \beta R(x)
        \right) \notag \\
        & \quad + O(n^2).
\end{align}

The remaining term is calculated as follows:
\begin{align}
    \frac{N}{2} \sum_{a,b} i\widetilde{Q}^{ab} Q^{ab}
    &= \frac{N}{2} \sum_{a,b} (\tilde{q} + \delta_{ab}(\widetilde{Q}-\tilde{q})) (q + \delta_{ab}(Q-q)) \notag \\
    &= \frac{Nn}{2} \widetilde{Q}Q - \frac{Nn}{2} \tilde{q}q + O(n^2), \\
    N \sum_{a=1}^n i\tilde{m}^a m^a
    &= N n \tilde{m} m.
\end{align}

Combining the expressions above yields
\begin{align}
    \mathbb{E}_{\bm{A}, \bm{x}^0} [Z^n]
    &=
        \int dQ d\widetilde{Q} dq d\tilde{q} dm d\tilde{m} \notag \\
    & \quad
        \exp \Biggl\{
            Nn \biggl(
            \frac{\widetilde{Q}Q}{2}
            - \frac{\tilde{q}q}{2}
            + \tilde{m}m
            - \beta \alpha \Psi_\text{RS}
            - \beta \widetilde{\Psi}_\text{RS}
            \biggr)
        \Biggr\},
\end{align}
where $\Psi_\text{RS}$ and $\widetilde{\Psi}_\text{RS}$ are defined in Eqs.~\eqref{eq:Psi-RS} and \eqref{eq:Psi-tilde-RS}, respectively.
Then, we apply $-\beta f_\text{RS} = \lim_{N \to \infty} N^{-1} \lim_{n \to 0^+} n^{-1} \log \mathbb E[Z^n]$.
Since the contribution to the integral is dominated by the saddle point as $N \to \infty$, we obtain Eq.~\eqref{eq:rs-free-energy}.

\section[Correspondence between the RS saddle point equations and the fixed point conditions of SE]%
{Correspondence between Eqs.~\eqref{eq:SPE-hat-Q}--\eqref{eq:RS-effective-field} and the fixed point conditions of Eqs.~\eqref{eq:SE-effective-field}--\eqref{eq:SE-chi}}%
\label{sec:apdx:rs-se-correspondence}
In this appendix, we show that the RS saddle point equations in Eqs.~\eqref{eq:SPE-hat-Q}--\eqref{eq:RS-effective-field} coincide with the fixed-point conditions of the SE recursion in Eqs.~\eqref{eq:SE-effective-field}--\eqref{eq:SE-chi}.
Notice that the MSE is defined by Eq.~\eqref{eq:replica-mse}.
At the RS saddle point, the order parameters $Q$ and $m$ are self-averaging, so this definition reduces to $\text{MSE}=Q-2m+\rho$.
Eqs.~\eqref{eq:RS-effective-estimator} and \eqref{eq:RS-effective-field} thus become
\begin{align}
    h_\text{RS} &= x^0 + \sqrt{\frac{\text{MSE}}{\alpha}}\,\xi, \\
    x^* &= S\!\left( h_\text{RS} ;\frac{\chi}{\alpha}R(\cdot) \right).
\end{align}
These expressions are identical to the scalar estimation rule in SE when its iteration-dependent quantities are replaced by their fixed-point values.
From here on, we need to show that the $\mathrm{MSE}$ and $\chi$ on the right-hand side of the above equations coincide with those of the fixed point of SE.

First, we show that the RS saddle-point equations for $Q$ and $m$ are equivalent to the fixed-point condition of Eq.~\eqref{eq:SE-mse}.
Eqs.~\eqref{eq:SPE-Q} and \eqref{eq:SPE-m}, together with $\mathbb{E}_{x^0}[(x^0)^2]=\rho$, give
\begin{align}
    \text{MSE}
        &= \mathbb{E}_{\bm{x},\bm{x}^0}\!\left[Q-2m\right]+\rho \notag \\
        &= \mathbb{E}_{x^0}\int D\xi\, \left[(x^*)^2-2x^0x^*+(x^0)^2\right] \notag \\
        &= \mathbb{E}_{x^0}\int D\xi\, \left(x^*-x^0\right)^2.
\end{align}
Thus, the RS equation for $\text{MSE}$ is the fixed-point condition of Eq.~\eqref{eq:SE-mse}.

Next, we establish the correspondence for $\chi$.
Its RS saddle-point equation reads
\begin{align}
    \chi = \frac{1}{\sqrt{\hat{\chi}}}
        \mathbb{E}_{x^0}\int D\xi\,\xi x^*.
\end{align}
For fixed $x^0$, define
\begin{align}
    f_{x^0}(\xi)
        = S\!\left(
            x^0+\frac{\sqrt{\hat{\chi}}}{\widehat{Q}}\xi;
            \frac{1}{\widehat{Q}}R(\cdot)
        \right).
\end{align}
By using Gaussian integration by parts, whose validity is established later, we obtain
\begin{align}
    \int D\xi\,\xi f_{x^0}(\xi)
        = \int D\xi\,f_{x^0}'(\xi).
\end{align}
Then, applying the chain rule almost everywhere yields
\begin{align}
    f_{x^0}'(\xi)
        = \frac{\sqrt{\hat{\chi}}}{\widehat{Q}}
        S'\!\left(h_\text{RS};\frac{1}{\widehat{Q}}R(\cdot)\right).
\end{align}
Substitution into the RS equation for $\chi$ gives
\begin{align}
    \chi
    &=
        \frac{1}{\widehat{Q}}\mathbb{E}_{x^0}\int D\xi\,
        S'\!\left(h_\text{RS};\frac{1}{\widehat{Q}}R(\cdot)\right) \notag \\
    &=
        \frac{\chi}{\alpha}\mathbb{E}_{x^0}\int D\xi\,
        S'\!\left(h_\text{RS};\frac{\chi}{\alpha}R(\cdot)\right).
\end{align}
This equation is identical to the fixed-point condition of Eq.~\eqref{eq:SE-chi}.

It remains to justify the Gaussian integration by parts used above.
Let $\lambda=1/\widehat{Q}$.
Formally, the following lemma holds:
\begin{lemma}
    If $\varepsilon^2\geq\lambda$, then $f_{x^0}(\xi) = S\!\left( x^0+\sqrt{\hat{\chi}} \lambda \xi; \lambda R(\cdot) \right)$ is locally absolutely continuous and satisfies
    \begin{align}
        \int D\xi\,\left|\xi f_{x^0}(\xi)\right| < \infty,
        \qquad
        \int D\xi\,\left|f_{x^0}'(\xi)\right| < \infty.
        \label{eq:lemma-goal-1}
    \end{align}
    Hence,
    \begin{align}
        \int D\xi\,\xi f_{x^0}(\xi)
        =
        \int D\xi\,f_{x^0}'(\xi).
        \label{eq:lemma-goal-2}
    \end{align}
\end{lemma}

\begin{proof}
    We first prove Eq.~\eqref{eq:lemma-goal-1} for $\varepsilon^2>\lambda$ and then treat the boundary case $\varepsilon^2=\lambda$ separately.
    \begin{itemize}
        \item \textit{Case $\varepsilon^2>\lambda$.}
        We first establish the local absolute continuity.
        For the $|x|>x_\text{sw}$ branch of Eq.~\eqref{eq:logsum-derivative}, the denominator satisfies
        \begin{align}
            (|z|+\varepsilon)^2-\lambda
            \geq
            \varepsilon^2-\lambda
            >0.
        \end{align}
        Hence, $S'$ is uniformly bounded.
        Since $S$ is continuous at the thresholds and continuously differentiable elsewhere, $S$ is locally absolutely continuous.
        Therefore, $f_{x^0}$ is also locally absolutely continuous, and the chain rule holds almost everywhere:
        \begin{align}
            f_{x^0}'(\xi)
            =
            \sqrt{\hat{\chi}}\lambda\,
            S'\!\left(
                x^0+\sqrt{\hat{\chi}}\lambda\xi;
                \lambda R(\cdot)
            \right).
        \end{align}

        We next establish the required absolute integrability.
        The stationarity condition gives
        \begin{align}
            |S(x;\lambda R(\cdot))|
            &=
            \begin{dcases}
                |x| - \frac{\lambda} {|S(x;\lambda R(\cdot))|+\varepsilon}
                    & \text{for } |x| > x_\text{sw}, \\
                0
                    & \text{for } |x| \le x_\text{sw}.
            \end{dcases}
        \end{align}
        This condition implies $|S(x;\lambda R(\cdot))| \leq |x|$ for all $x$, and consequently
        \begin{align}
            |\xi f_{x^0}(\xi)| \leq |\xi| |S(x^0+\sqrt{\hat{\chi}} \lambda \xi;\lambda R(\cdot))| \leq |x^0||\xi| + \sqrt{\hat{\chi}}\lambda |\xi|^2.
        \end{align}
        Since the Gaussian measure has finite first and second moments,
        \begin{align}
            \int D\xi\,
            \left|\xi f_{x^0}(\xi)\right|
            <\infty.
        \end{align}
        Moreover, the uniform boundedness of $S'$ for $\varepsilon^2>\lambda$ directly implies
        \begin{align}
            \int D\xi\,|f_{x^0}'(\xi)|<\infty.
        \end{align}

        \item \textit{Boundary case $\varepsilon^2=\lambda$.}
        As in the case $\varepsilon^2>\lambda$, we verify the local absolute continuity of $f_{x^0}$ and the required absolute integrability.

        We begin with the behavior near the thresholds $|x|=x_\text{sw}=\varepsilon$.
        For $x>\varepsilon$, let $z=S(x;\varepsilon^2R(\cdot))>0$.
        The stationarity condition gives
        \begin{align}
            x - \varepsilon = \frac{z^2}{z+\varepsilon} \quad \text{for } x > \varepsilon.
        \end{align}
        Hence, $z \sim \sqrt{\varepsilon (x - \varepsilon)}$ as $x\to\varepsilon^+$.
        Substitution into Eq.~\eqref{eq:logsum-derivative} yields
        \begin{align}
            S'(x;\varepsilon^2R(\cdot))
            \sim
            \frac{1}{2}\sqrt{\frac{\varepsilon}{x-\varepsilon}}.
        \end{align}
        By symmetry, the same behavior occurs at the negative threshold.
        Thus, on the active side of either threshold,
        \begin{align}
            S'(x;\varepsilon^2R(\cdot))
            =
            O\!\left(
                \left||x|-\varepsilon\right|^{-1/2}
            \right).
        \end{align}
        This inverse-square-root singularity is still locally integrable since $\int_{+0}^\delta u^{-1/2}\,du = 2\sqrt{\delta}<\infty$.
        Together with the continuity of $S$ at the thresholds $|x| = \varepsilon$ and its continuous differentiability away from them, this shows that $S$ is locally absolutely continuous also for $\varepsilon^2=\lambda$.
        Accordingly, $f_{x^0}$ is locally absolutely continuous and satisfies the same almost-everywhere chain rule.

        We next establish the required absolute integrability.
        The growth bound for $\xi f_{x^0}(\xi)$ derived above is also valid for $\varepsilon^2 = \lambda$.
        Furthermore, since the argument of $S$ is affine in $\xi$, each threshold $x = \pm \varepsilon$ corresponds to at most one point $\xi = \xi_{\pm}$.
        Near such points, $|x \mp \varepsilon| \propto |\xi - \xi_{\pm}|$, which implies the inverse-square-root singularity of $S'$ remains an inverse-square-root singularity in $f'_{x^0}$, and is therefore locally integrable.
        Since $S'(x)\to1$ as $|x|\to\infty$, we obtain
        \begin{align}
            \int D\xi\,|f_{x^0}'(\xi)|<\infty.
        \end{align}
    \end{itemize}

    Next, we prove Eq.~\eqref{eq:lemma-goal-2}.
    Let $\phi(\xi)=e^{-\xi^2/2}/\sqrt{2\pi}$.
    Since $f_{x^0}$ is absolutely continuous on every finite interval, integration by parts gives
    \begin{align}
        \int_{-L}^{L}
        f_{x^0}'(\xi)\phi(\xi)\,d\xi
        =
        \left[
            f_{x^0}(\xi)\phi(\xi)
        \right]_{-L}^{L}
        +
        \int_{-L}^{L}
        \xi f_{x^0}(\xi)\phi(\xi)\,d\xi.
    \end{align}
    The linear-growth bound on $f_{x^0}$ implies
    \begin{align}
        f_{x^0}(\xi)\phi(\xi)\to0
        \qquad
        (|\xi|\to\infty).
    \end{align}
    Taking $L\to\infty$ and using the absolute integrability established above yields
    \begin{align}
        \int D\xi\,f_{x^0}'(\xi)
        =
        \int D\xi\,\xi f_{x^0}(\xi).
    \end{align}
\end{proof}

\section[Derivation of linear stability condition]{Derivation of Eqs.~\eqref{eq:self-consistent-hat-chi} and \eqref{eq:alpha_c-log-sum}}
\label{sec:apdx:linear-stability}

In this appendix, we derive the single self-consistent equation for $\hat{\chi}$ given in Eq.~\eqref{eq:self-consistent-hat-chi}, which describes the system in the limit of exact recovery, i.e., $Q \to \rho$, $m \to \rho$, $\chi \to 0^+$, $\widehat{Q} \to \infty$, and $\hat{m} \to \infty$.
We also derive the linear stability condition for this fixed point.

We start from the following equation obtained by removing $\chi$ from Eq.~\eqref{eq:SPE-hat-Q} and Eq.~\eqref{eq:SPE-hat-chi}:
\begin{align}
    \hat{\chi}
        &= \frac{\widehat{Q}^2 (Q - 2m + \rho)}{\alpha} \notag \\
        &= \frac{\widehat{Q}^2}{\alpha} \mathbb{E}_{x^0} \int D\xi (x^* - x^0)^2.
\end{align}
This equation suggests that we need to evaluate the asymptotic forms of $x^* - x^0$ up to the order of $O(\widehat{Q}^{-1})$ to derive the self-consistent equation for $\hat{\chi}$.
Thus, we first analyze the asymptotic behavior of the estimator $x^*$ defined in Eq.~\eqref{eq:RS-effective-estimator}, and then use the result to evaluate $x^* - x^0$.

Hereafter, we assume that $\varepsilon$ satisfies the convexity condition $\varepsilon \ge \widehat{Q}^{-1/2}$.
When this condition is satisfied, $x^*$ is given by the following expression
\begin{align}
    x^*
    &=
        \begin{dcases}
            \sign\!\left(
                h_\text{RS}
            \right)
            \left(
                \frac{ \left| h_\text{RS} \right| - \varepsilon }{2}
                + \sqrt{
                    \frac{ (\left| h_\text{RS} \right| + \varepsilon)^2 }{4} - \frac{1}{\widehat{Q}}
                }
            \right)
            & \text{if } \left| h_\text{RS} \right| > \frac{1}{\widehat{Q} \varepsilon},
            \\
            0 & \text{otherwise},
        \end{dcases}
\end{align}
where $h_\text{RS} = x^0 + (\sqrt{\hat{\chi}} / \widehat{Q}) \xi$.
Here, we consider two cases separately:~when $x^0 = 0$ and $x^0 \sim \mathcal{N}(0, 1)$.
\begin{itemize}
\item
For $x^0 = 0$, the expression becomes
\begin{align}
    x^*_{x^0 = 0}
    &=
        \begin{dcases}
            \sign\!\left(
                \xi
            \right)
            \left(
                \frac{1}{2} \left( \left| \frac{ \sqrt{\hat{\chi}} }{ \widehat{Q} } \xi \right| - \varepsilon \right)
                + \sqrt{
                    \frac{1}{4} \left( \left| \frac{ \sqrt{\hat{\chi}} }{ \widehat{Q} } \xi \right| + \varepsilon \right)^2 - \frac{1}{\widehat{Q}}
                }
            \right)
            & \text{if } |\xi| > \frac{1}{\sqrt{\varepsilon^2 \hat{\chi}}},
            \\
            0 & \text{otherwise}.
        \end{dcases}
\end{align}
Taking a first-order expansion around $\widehat{Q}^{-1} = 0$, we obtain
\begin{align}
    x^*_{x^0 = 0} =
    \begin{dcases}
        \frac{1}{\widehat{Q}} \left( \sqrt{\hat{\chi}} \xi - \frac{\sign(\xi)}{\varepsilon} \right) + O(\widehat{Q}^{-2})
            & \text{if } |\xi| > \frac{1}{\sqrt{\varepsilon^2 \hat{\chi}}}, \\
        0
            & \text{otherwise}.
    \end{dcases}
\end{align}
The contribution from this part is therefore
\begin{align}
    \widehat{Q}^2 (1 - \rho) \int D\xi (x^*_{x^0 = 0})^2
    &=
        (1 - \rho) \int_{|\xi| > 1 / \sqrt{\varepsilon^2 \hat{\chi}} } D\xi\, \left( \sqrt{ \hat{\chi} } \xi - \frac{\sign(\xi)}{\varepsilon}\right)^2 \notag \\
    &=
        \frac{2 (1 - \rho)}{\varepsilon^2} \left(
            \left( \varepsilon^2 \hat{\chi} + 1 \right) H \!\left( \frac{1}{ \sqrt{\varepsilon^2 \hat{\chi}} } \right)
            - \sqrt{ \varepsilon^2 \hat{\chi} } \phi \!\left( \frac{1}{ \sqrt{\varepsilon^2 \hat{\chi}} } \right)
        \right).
    \label{eq:hatchi-contrib-zero}
\end{align}

\item
For $x^0 \neq 0$, taking a first-order expansion around $\widehat{Q}^{-1} = 0$ yields
\begin{align}
    x^*_{x^0 \neq 0} =
    \begin{dcases}
        x^0 + \frac{1}{\widehat{Q}} \left( \sqrt{\hat{\chi}} \xi - \frac{\sign(x^0)}{|x^0| + \varepsilon} \right) + O(\widehat{Q}^{-2})
        & \text{if } \left|\widehat{Q} x^0 + \sqrt{\hat{\chi}} \xi\right| > \frac{1}{\varepsilon}, \\
        0 & \text{otherwise}.
    \end{dcases}
\end{align}
The contribution from this part is given by
\begin{align}
    \widehat{Q}^2 \rho \int Dx^0 \int D\xi (x^*_{x^0 \neq 0} - x^0)^2
    &= \rho \int Dx^0 \int D\xi \left( \sqrt{\hat{\chi}} \xi - \frac{\sign(x^0)}{|x^0| + \varepsilon} \right)^2 \notag \\
    &= \rho \left( \hat{\chi} + \int D\xi \frac{1}{\left(|\xi| + \varepsilon\right)^2} \right).
    \label{eq:hatchi-contrib-nonzero}
\end{align}
Here, the contribution from the case $x^*_{x^0 \neq 0} = 0$ is of order $\widehat{Q}^{-1}$ and is thus neglected.
\end{itemize}
By summing the contributions from Eqs.~\eqref{eq:hatchi-contrib-zero} and \eqref{eq:hatchi-contrib-nonzero}, the reduced self-consistent equation for $\hat{\chi}$ is derived as Eq.~\eqref{eq:self-consistent-hat-chi}.

For the fixed point $\hat{\chi}^*$ to be locally stable, $|F'(\hat{\chi}^*)| < 1$ should hold, where $F(\hat{\chi})$ is the right-hand side of Eq.~\eqref{eq:self-consistent-hat-chi} divided by $\varepsilon^2$.
Differentiating $F(\hat{\chi})$ and using the identities $\phi'(z) = -z\phi(z)$ and $H'(z) = -\phi(z)$ leads to the following expression:
\begin{align}
    |F'(\hat{\chi})| = \frac{1}{\alpha} \left(
        2(1 - \rho) H\!\left( \frac{1}{\sqrt{\varepsilon^2 \hat{\chi}}} \right) + \rho
    \right).
\end{align}
Thus, the stability condition is given as $\alpha > \alpha_c$, where $\alpha_c$ is defined in Eq.~\eqref{eq:alpha_c-log-sum}.

\section[Derivation of dAT stability condition]%
{Derivation of Eq.~\eqref{eq:dat-condition}}%
\label{sec:apdx:1rsb-free-energy}

This appendix explains the derivation of the dAT condition.
To derive this condition, we first formulate the saddle-point equations under the one-step replica symmetry breaking (1RSB) ansatz, which is a one-step generalization of the RS ansatz.
We then determine the condition under which the system returns to the RS fixed point when a small perturbation is applied in a symmetry-breaking direction.

\subsection{1RSB free energy density}

We consider the following block structure of the order parameters.
The replica indices $a, b \in \{1, \dots, n\}$ are divided into $n/m_1$ blocks of size $m_1$.
Each replica index $a$ is uniquely represented using a block index $a_0 \in \{1, \dots, n/m_1\}$ and an intra-block index $a_1 \in \{1, \dots, m_1\}$ as $a = m_1(a_0 - 1) + a_1$.
Based on this hierarchical structure, the order parameters are parameterized as follows:
\begin{align}
Q^{ab} &= Q \delta_{ab} + q_1 (\delta_{a_0 b_0} - \delta_{ab}) + q (1 - \delta_{a_0 b_0}), \\
i\widetilde{Q}^{ab} &= \widetilde{Q} \delta_{ab} + \tilde{q}_1 (\delta_{a_0 b_0} - \delta_{ab}) + \tilde{q} (1 - \delta_{a_0 b_0}), \\
m^a &= m, \\
i\tilde{m}^a &= \tilde{m}.
\end{align}

Under this ansatz, the free energy density is simplified as follows.
Similarly to the derivation of the RS solution in Appendix \ref{sec:apdx:rs-free-energy}, we hierarchically decompose the random variables $y^a - y^0$ by introducing $1 + n/m_1 + n$ standard Gaussian variables $\xi_0, \{\xi_1^{a_0}\}_{a_0}, \{\xi_2^{a_0, a_1}\}_{a_0, a_1}$ as
\begin{gather}
y^a - y^0 = \sqrt{E} \, \xi_0 + \sqrt{\Delta_1} \, \xi_1^{a_0} + \sqrt{\Delta_2} \, \xi_2^{a_0, a_1},
\end{gather}
where
\begin{align}
    E &= q - 2m + \rho, \\
    \Delta_1 &= q_1 - q, \\
    \Delta_2 &= Q - q_1.
\end{align}
This decomposition and Gaussian integration lead to the following expression:
\begin{align}
    n \beta \Psi
    =
    \frac{n}{2} \left(
        \frac{E}{\sigma^2 + \Delta_2 + m_1 \Delta_1}
        + \frac{1}{m_1} \log \!\left( 1 + \frac{m_1 \Delta_1}{\sigma^2 + \Delta_2} \right)
        + \log \!\left( 1 + \frac{\Delta_2}{\sigma^2} \right)
    \right) + O(n^2).
\end{align}
For the term $\widetilde{\Psi}$, we linearize the quadratic interaction $\sum_{a,b} i\widetilde{Q}^{ab} x^a x^b$ by applying the Hubbard-Stratonovich transformation.
The single-site integral is then simplified as
\begin{align}
    n \beta \widetilde{\Psi}
    &=
    -\log \mathbb{E}_{x^0} \int D\xi_0 \left(
        \int D\xi_1 \left(
            \int dx \exp \!\left(
                -\beta \mathcal{H}_{\text{eff}}(x, \xi_0, \xi_1, x^0)
            \right)
        \right)^{m_1}
    \right)^{n/m_1},
\end{align}
where
\begin{align}
    \beta \mathcal{H}_{\text{eff}}(x, \xi_0, \xi_1, x^0) = \frac{\widetilde{Q} - \tilde{q}_1}{2} x^2 - (\sqrt{-\tilde{q}} \xi_0 + \sqrt{-(\tilde{q}_1 - \tilde{q})} \xi_1 - \tilde{m} x^0)x + \beta R(x).
\end{align}
The remaining term is simplified as
\begin{align}
\frac{N}{2} \sum_{a,b} i\widetilde{Q}^{ab} Q^{ab}
&= \frac{Nn}{2} \left( \widetilde{Q} Q + (m_1 - 1) \tilde{q}_1 q_1 + (n - m_1) \tilde{q} q \right).
\end{align}
By combining these terms, taking the limit $n \to 0^+$ and applying the saddle-point method, the free energy density is obtained as the saddle-point condition with respect to the order parameters $\Theta = \{Q, q_1, q, m, \widetilde{Q}, \tilde{q}_1, \tilde{q}, \tilde{m}\}$ as follows:
\begin{align}
    -\beta f_\text{1RSB} = \operatorname*{extr}_{\Theta} \!\left\{
        \frac{\widetilde{Q} Q}{2} - \frac{(1 - m_1) \tilde{q}_1 q_1}{2} - \frac{m_1 \tilde{q} q}{2} + \tilde{m} m - \beta \alpha \Psi_\text{1RSB} - \beta \widetilde{\Psi}_\text{1RSB}
    \right\},
\end{align}
where
\begin{align}
    \beta \Psi_\text{1RSB}
    &=
        \frac{1}{2} \Bigg(
            \frac{ E }{ \sigma^2 + \Delta_2 + m_1 \Delta_1 }
            + \frac{1}{m_1} \log\!\left(\sigma^2 + \Delta_2 + m_1 \Delta_1\right)
            - \frac{1 - m_1}{m_1} \log \!\left( \sigma^2 + \Delta_2 \right)
            \notag \\ & \qquad
            - \log \sigma^2
        \Bigg), \\
    \beta \widetilde{\Psi}_\text{1RSB}
    &=
        -\frac{1}{m_1} \mathbb{E}_{x^0} \int D\xi_0 \log \int D\xi_1 \left(
            \int dx \exp \!\left(
                -\beta \mathcal{H}_{\text{eff}}(x, \xi_0, \xi_1, x^0)
            \right)
        \right)^{m_1}.
\end{align}
If $m_1 = 1$, this saddle-point condition reduces to the RS solution.

\subsection{dAT stability condition}

To obtain the dAT stability condition, we fix the order parameters other than $q_1$ and $\tilde{q}_1$ at their RS values, denoted by $Q_\text{RS}$, $q_\text{RS}$, $m_\text{RS}$, $\widetilde{Q}_\text{RS}$, $\tilde{q}_\text{RS}$ and $\tilde{m}_\text{RS}$, and derive the asymptotic self-consistent equation of $q_1 - q_\text{RS}$ when $q_1 \approx q_\text{RS}$.
By deriving the condition that $q_1 - q_\text{RS}$ does not expand per iteration, the AT stability condition can be determined.
Here, we use the following scaling of the RS order parameters:
\begin{align}
    \chi_\text{RS} &= \beta (Q_\text{RS} - q_\text{RS}), \\
    \widehat{Q}_\text{RS} &= \frac{\widetilde{Q}_\text{RS} - \tilde{q}_\text{RS}}{\beta}, \\
    \hat{\chi}_\text{RS} &= -\frac{\tilde{q}_\text{RS}}{\beta^2}.
\end{align}
The deviations from the RS solution are explained by the following variables:
\begin{align}
    \widehat{\Delta} &= -\frac{\tilde{q}_1 - \tilde{q}_\text{RS}}{\beta^2}, \\
    \Delta &= q_1 - q_\text{RS}.
\end{align}
If the deviations $\widehat{\Delta}$ and $\Delta$ expand, the RS solution is unstable.

First, we derive the saddle-point equation for $\widehat{\Delta}$.
From the self-consistent equations of $\tilde{q}$ and $\tilde{q}_1$ obtained by differentiating $f_\text{1RSB}$ with respect to $q$ and $q_1$, respectively, the equation for $\widehat{\Delta}$ is obtained as follows:
\begin{align}
    \widehat{\Delta} = \frac{\alpha}{\chi_\text{RS}^2} \Delta + O(\Delta^2).
    \label{eq:1rsb-delta<-hatdelta}
\end{align}

Next, we derive the saddle-point equation for $\Delta$.
We differentiate $f_\text{1RSB}$ with respect to $\tilde{q}$ and $\tilde{q}_1$, respectively, to obtain the equations for $q$ and $q_1$.
By calculating their difference, $\Delta$ can be interpreted as the variance of the estimator in the effective single-site problem with respect to the intra-block auxiliary field $\xi_1$:
\begin{align}
    \Delta = \mathbb{E}_{x^0} \int D\xi_0 \left( \left\langle x^2 \right\rangle_{\xi_1} - \left\langle x \right\rangle_{\xi_1}^2 \right).
\end{align}
Here, $\langle \cdot \rangle_{\xi_1}$ denotes the average with respect to the following probability density function $\pi_1(\xi_1 \mid \xi_0, x^0)$:
\begin{align}
    \pi_1(\xi_1 \mid \xi_0, x^0)
    = \frac{1}{\mathcal{Z}(\xi_0, x^0)} \exp\!\left( -\frac{\xi_1^2}{2} \right) \Biggl( \int dx \exp \Bigl( -\beta \mathcal{H}_{\text{eff}}(x, \xi_0, \xi_1, x^0) \Bigr) \Biggr)^{m_1}.
\end{align}
In the low-temperature limit $\beta \to \infty$, $\pi_1$ is dominated by the point $x=x^*$ that minimizes $\mathcal{H}_{\text{eff}}$, and the integral concentrates around $x^*$.
In the vicinity of the RS solution, the estimator $x^*$ is expressed using the thresholding function as follows:
\begin{align}
    x^* = S\!\left(
        h_\text{1RSB}; \frac{1}{\widehat{Q}_\text{RS}} R(\cdot)
    \right).
\end{align}
Here,
\begin{align}
    h_\text{1RSB}
    &= h_\text{RS} + \frac{\sqrt{\widehat{\Delta}}}{\widehat{Q}_\text{RS}} \xi_1
    =
        x^0
        + \frac{\sqrt{\hat{\chi}_\text{RS}}}{\widehat{Q}_\text{RS}} \xi_0
        + \frac{\sqrt{\widehat{\Delta}}}{\widehat{Q}_\text{RS}} \xi_1,
\end{align}
and the deviation from the RS solution is given by $\sqrt{\widehat{\Delta}}\, \xi_1 / \widehat{Q}_\text{RS}$.
Regarding this quantity as infinitesimal, we expand $x^*$ around $x^*_\text{RS}$ as
\begin{align}
    x^*
    =
        x^*_\text{RS}
        + \frac{\partial x^*}{\partial h} \bigg|_{ x^* = x^*_\text{RS} } \frac{\sqrt{\widehat{\Delta}}}{\widehat{Q}_\text{RS}} \xi_1
        + O(\widehat{\Delta}),
\end{align}
where
\begin{align}
    x^*_\text{RS}
    &=
        S\!\left(
            h_\text{RS};
            \frac{1}{\widehat{Q}_\text{RS}} R(\cdot)
        \right), \\
    \frac{\partial x^*}{\partial h} \bigg|_{ x^* = x^*_\text{RS} }
    &=
        S'\!\left(
            h_\text{RS};
            \frac{1}{\widehat{Q}_\text{RS}} R(\cdot)
        \right).
\end{align}
Using this expansion, the variance term is evaluated as
\begin{align}
    \langle x^2 \rangle_{\xi_1} - \langle x \rangle_{\xi_1}^2
    &\simeq
        \frac{1}{\widehat{Q}_\text{RS}^2}
        S'\!\left( h_\text{RS} ; \frac{1}{\widehat{Q}_\text{RS}} R(\cdot) \right)^2
        \widehat{\Delta} \times \sigma_{\xi_1}^2,
\end{align}
where $\sigma_{\xi_1}^2$ is the variance of $\xi_1 \sim \pi_1(\xi_1 \mid \xi_0, x^0)$.
The distribution $\pi_1$ is expressed as $\pi_1(\xi_1 \mid \xi_0, x^0) \propto \exp(-\phi(\xi_1))$ using the effective potential $\phi(\xi_1) \coloneqq (1/2) \xi_1^2 + m_1 \beta \mathcal{H}_{\text{eff}}(x^*, \xi_1)$.
The variance $\sigma_{\xi_1}^2$ can be approximated by the inverse of the curvature of the potential $\phi''(\xi_1)^{-1}$, where the curvature $\phi''(\xi_1)$ is evaluated as follows:
\begin{align}
    \phi''(\xi_1)
    =
    1 + m_1 \beta \left( \frac{\partial^2 \mathcal{H}_{\text{eff}}}{\partial \xi_1^2} + \frac{\partial^2 \mathcal{H}_{\text{eff}}}{\partial x \partial \xi_1} \frac{\partial x^*}{\partial \xi_1} \right) = 1 + O(\widehat{\Delta}).
\end{align}
These results lead to the following relationship:
\begin{align}
    \Delta
    =
    \frac{1}{\widehat{Q}_\text{RS}^2}
    \mathbb{E}_{x^0} \int D\xi
    S'\!\left( h_\text{RS} ; \frac{1}{\widehat{Q}_\text{RS}} R(\cdot) \right)^2
    \widehat{\Delta}
    + O(\widehat{\Delta}^2).
    \label{eq:1rsb-hatdelta<-delta}
\end{align}

Combining Eqs.~\eqref{eq:1rsb-delta<-hatdelta} and \eqref{eq:1rsb-hatdelta<-delta}, the condition for the perturbation $\Delta$ to contract to zero is derived as follows:
\begin{align}
    \frac{\alpha}{\chi_\text{RS}^2}
    \frac{1}{\widehat{Q}_\text{RS}^2}
    \mathbb{E}_{x^0} \int D\xi
    S'\!\left( h_\text{RS} ; \frac{1}{\widehat{Q}_\text{RS}} R(\cdot) \right)^2
    < 1.
\end{align}
Applying $\widehat{Q}_\text{RS} = \alpha / \chi_\text{RS}$ and $h_\text{RS} = x^0 + \sqrt{\hat{\chi}_\text{RS}} \xi / \widehat{Q}_\text{RS}$ to this expression reduces it to Eq.~\eqref{eq:dat-condition}.

\bibliography{references}

@ARTICLE{Bhadra2019-il,
  title     = "Lasso meets horseshoe: A survey",
  author    = "Bhadra, Anindya and Datta, Jyotishka and Polson, Nicholas G and
               Willard, Brandon",
  journal   = "Stat. Sci.",
  publisher = "Institute of Mathematical Statistics",
  volume    =  34,
  number    =  3,
  pages     = "405--427",
  month     =  "1~" # aug,
  year      =  2019,
  url       = "http://dx.doi.org/10.1214/19-sts700"
}

@BOOK{Foucart2013-ej,
  title     = "A Mathematical Introduction to Compressive Sensing",
  author    = "Foucart, Simon and Rauhut, Holger",
  publisher = "Birkhauser Boston",
  address   = "Secaucus, NJ",
  edition   =  2013,
  series    = "Applied and Numerical Harmonic Analysis",
  month     =  "8~" # aug,
  year      =  2013,
  url       = "http://dx.doi.org/10.1007/978-0-8176-4948-7"
}

@ARTICLE{Malek-Mohammadi2016-ix,
  title     = "Successive concave sparsity approximation for compressed sensing",
  author    = "Malek-Mohammadi, Mohammadreza and Koochakzadeh, Ali and
               Babaie-Zadeh, Massoud and Jansson, Magnus and Rojas, Cristian R",
  journal   = "IEEE Trans. Signal Process.",
  publisher = "Institute of Electrical and Electronics Engineers (IEEE)",
  volume    =  64,
  number    =  21,
  pages     = "5657--5671",
  month     =  "1~" # nov,
  year      =  2016,
  url       = "http://dx.doi.org/10.1109/tsp.2016.2585096"
}

@ARTICLE{Hadj-Selem2018-zk,
  title     = "Continuation of Nesterov's smoothing for regression with
               structured sparsity in high-dimensional neuroimaging",
  author    = "Hadj-Selem, Fouad and Lofstedt, Tommy and Dohmatob, Elvis and
               Frouin, Vincent and Dubois, Mathieu and Guillemot, Vincent and
               Duchesnay, Edouard",
  journal   = "IEEE Trans. Med. Imaging",
  publisher = "Institute of Electrical and Electronics Engineers (IEEE)",
  volume    =  37,
  number    =  11,
  pages     = "2403--2413",
  month     =  nov,
  year      =  2018,
  url       = "http://dx.doi.org/10.1109/tmi.2018.2829802⟩"
}

@INPROCEEDINGS{Donoho2010-kk,
  title     = "Message passing algorithms for compressed sensing: {II}. analysis
               and validation",
  author    = "Donoho, David L and Maleki, Arian and Montanari, Andrea",
  booktitle = "IEEE Information Theory Workshop 2010 (ITW 2010)",
  publisher = "IEEE",
  pages     = "1--5",
  month     =  jan,
  year      =  2010,
  url       = "http://dx.doi.org/10.1109/ITWKSPS.2010.5503228"
}

@ARTICLE{Barbier2025-ob,
  title     = "Phase diagram of compressed sensing with ℓ 0 -norm regularization",
  author    = "Barbier, Damien and Lucibello, Carlo and Saglietti, Luca and
               Krzakala, Florent and Zdeborová, Lenka",
  journal   = "J. Stat. Mech.",
  publisher = "IOP Publishing",
  volume    =  2025,
  number    =  6,
  pages     =  063402,
  month     =  "1~" # jun,
  year      =  2025,
  url       = "http://dx.doi.org/10.1088/1742-5468/adca21"
}

@INPROCEEDINGS{Donoho2010-mz,
  title     = "Message passing algorithms for compressed sensing: {I}.
               motivation and construction",
  author    = "Donoho, David L and Maleki, Arian and Montanari, Andrea",
  booktitle = "IEEE Information Theory Workshop 2010 (ITW 2010)",
  publisher = "IEEE",
  pages     = "1--5",
  month     =  jan,
  year      =  2010,
  url       = "http://dx.doi.org/10.1109/ITWKSPS.2010.5503193"
}

@ARTICLE{Armagan2013-js,
  title     = "Generalized double Pareto shrinkage",
  author    = "Armagan, Artin and Dunson, David B and Lee, Jaeyong",
  journal   = "Stat. Sin.",
  publisher = "Institute of Statistical Science",
  volume    =  23,
  number    =  1,
  pages     = "119--143",
  month     =  "1~" # jan,
  year      =  2013,
  url       = "http://dx.doi.org/10.5705/ss.2011.048"
}

@ARTICLE{Wu2012-oe,
  title     = "Optimal Phase Transitions in Compressed Sensing",
  author    = "Wu, Yihong and Verdu, Sergio",
  journal   = "IEEE Trans. Inf. Theory",
  publisher = "Institute of Electrical and Electronics Engineers (IEEE)",
  volume    =  58,
  number    =  10,
  pages     = "6241--6263",
  month     =  oct,
  year      =  2012,
  url       = "http://dx.doi.org/10.1109/tit.2012.2205894"
}

@ARTICLE{Mohimani2009-lg,
  title   = "A Fast Approach for Overcomplete Sparse Decomposition Based on
             Smoothed $\ell ^0$ Norm",
  author  = "Mohimani, Hosein and Babaie-Zadeh, Massoud and Jutten, Christian",
  journal = "IEEE Transactions on Signal Processing",
  volume  =  57,
  number  =  1,
  pages   = "289--301",
  year    =  2009,
  url     = "http://dx.doi.org/10.1109/TSP.2008.2007606"
}

@inproceedings{Furuhashi2026WEEP,
  title = {{{WEEP}}: {{A Differentiable Nonconvex Sparse Regularizer}} via {{Weakly-Convex Envelope}}},
  shorttitle = {{{WEEP}}},
  booktitle = {2026 {{IEEE International Conference}} on {{Acoustics}}, {{Speech}} and {{Signal Processing}} ({{ICASSP}})},
  author = {Furuhashi, Takanobu and Hontani, Hidekata and Zhao, Qibin and Yokota, Tatsuya},
  year = 2026,
  month = may,
  pages = {56--60},
  publisher = {IEEE},
  issn = {2379-190X},
  doi = {10.1109/ICASSP55912.2026.11462436},
  url = {https://ieeexplore.ieee.org/abstract/document/11462436},
  urldate = {2026-09-03},
}

@ARTICLE{Ma2017-tk,
  title     = "Orthogonal {AMP}",
  author    = "Ma, Junjie and Ping, Li",
  journal   = "IEEE Access",
  publisher = "Institute of Electrical and Electronics Engineers (IEEE)",
  volume    =  5,
  pages     = "2020--2033",
  year      =  2017,
  url       = "http://dx.doi.org/10.1109/ACCESS.2017.2653119"
}

@ARTICLE{Ayach2014-qg,
  title     = "Spatially sparse precoding in millimeter wave {MIMO} systems",
  author    = "Ayach, Omar El and Rajagopal, Sridhar and Abu-Surra, Shadi and
               Pi, Zhouyue and Heath, Robert W",
  journal   = "IEEE Trans. Wirel. Commun.",
  publisher = "Institute of Electrical and Electronics Engineers (IEEE)",
  volume    =  13,
  number    =  3,
  pages     = "1499--1513",
  month     =  mar,
  year      =  2014,
  url       = "http://dx.doi.org/10.1109/twc.2014.011714.130846"
}

@ARTICLE{Wagadarikar2008-ef,
  title     = "Single disperser design for coded aperture snapshot spectral
               imaging",
  author    = "Wagadarikar, Ashwin and John, Renu and Willett, Rebecca and
               Brady, David",
  journal   = "Appl. Opt.",
  publisher = "Optica Publishing Group",
  volume    =  47,
  number    =  10,
  pages     = "B44--51",
  month     =  "1~" # apr,
  year      =  2008,
  url       = "http://dx.doi.org/10.1364/ao.47.000b44"
}

@ARTICLE{Duarte2008-sm,
  title     = "Single-pixel imaging via compressive sampling",
  author    = "Duarte, Marco F and Davenport, Mark A and Takhar, Dharmpal and
               Laska, Jason N and Sun, Ting and Kelly, Kevin F and Baraniuk,
               Richard G",
  journal   = "IEEE Signal Process. Mag.",
  publisher = "Institute of Electrical and Electronics Engineers (IEEE)",
  volume    =  25,
  number    =  2,
  pages     = "83--91",
  month     =  mar,
  year      =  2008,
  url       = "http://dx.doi.org/10.1109/msp.2007.914730"
}

@ARTICLE{Lustig2007-zw,
  title     = "Sparse {MRI}: The application of compressed sensing for rapid
               {MR} imaging",
  author    = "Lustig, Michael and Donoho, David and Pauly, John M",
  journal   = "Magn. Reson. Med.",
  publisher = "Wiley",
  volume    =  58,
  number    =  6,
  pages     = "1182--1195",
  month     =  dec,
  year      =  2007,
  url       = "http://dx.doi.org/10.1002/mrm.21391"
}

@ARTICLE{Coifman1992-mi,
  title     = "Entropy-based algorithms for best basis selection",
  author    = "Coifman, R R and Wickerhauser, M V",
  journal   = "IEEE Trans. Inf. Theory",
  publisher = "Institute of Electrical and Electronics Engineers (IEEE)",
  volume    =  38,
  number    =  2,
  pages     = "713--718",
  month     =  mar,
  year      =  1992,
  url       = "http://dx.doi.org/10.1109/18.119732"
}

@ARTICLE{Gu2025-wr,
  title         = "Perfect reconstruction of sparse signals using nonconvexity
                   control and one-step {RSB} message passing",
  author        = "Gu, Xiaosi and Sakata, Ayaka and Obuchi, Tomoyuki",
  journal       = "arXiv [stat.ML]",
  month         =  "19~" # dec,
  year          =  2025,
  url           = "http://arxiv.org/abs/2512.17426",
  archivePrefix = "arXiv",
  primaryClass  = "stat.ML"
}

@ARTICLE{Krzakala2012-is,
  title     = "Statistical-physics-based reconstruction in compressed sensing",
  author    = "Krzakala, F and Mézard, M and Sausset, F and Sun, Y F and
               Zdeborová, L",
  journal   = "Phys. Rev. X.",
  publisher = "American Physical Society (APS)",
  volume    =  2,
  number    =  2,
  pages     =  021005,
  month     =  "11~" # may,
  year      =  2012,
  url       = "http://dx.doi.org/10.1103/PhysRevX.2.021005"
}

@INPROCEEDINGS{Rangan2017-vh,
  title     = "Vector approximate message passing",
  author    = "Rangan, Sundeep and Schniter, Philip and Fletcher, Alyson K",
  booktitle = "2017 IEEE International Symposium on Information Theory (ISIT)",
  publisher = "IEEE",
  pages     = "1588--1592",
  month     =  jun,
  year      =  2017,
  url       = "https://dl.acm.org/doi/10.1109/ISIT.2017.8006797"
}

@ARTICLE{Bayati2011-de,
  title     = "The dynamics of message passing on dense graphs, with
               applications to compressed sensing",
  author    = "Bayati, Mohsen and Montanari, Andrea",
  journal   = "IEEE Trans. Inf. Theory",
  publisher = "Institute of Electrical and Electronics Engineers (IEEE)",
  volume    =  57,
  number    =  2,
  pages     = "764--785",
  month     =  feb,
  year      =  2011,
  url       = "http://dx.doi.org/10.1109/TIT.2010.2094817"
}

@ARTICLE{Sakata2018-ey,
  title     = "Approximate message passing for nonconvex sparse regularization
               with stability and asymptotic analysis",
  author    = "Sakata, Ayaka and Xu, Yingying",
  journal   = "J. Stat. Mech.",
  publisher = "IOP Publishing",
  volume    =  2018,
  number    =  3,
  pages     =  033404,
  month     =  "13~" # mar,
  year      =  2018,
  url       = "http://dx.doi.org/10.1088/1742-5468/aab051"
}

@ARTICLE{Zhang2010-gb,
  title     = "Nearly unbiased variable selection under minimax concave penalty",
  author    = "Zhang, Cun-Hui",
  journal   = "Ann. Stat.",
  publisher = "Institute of Mathematical Statistics",
  volume    =  38,
  number    =  2,
  pages     = "894--942",
  month     =  "1~" # apr,
  year      =  2010,
  url       = "https://projecteuclid.org/journals/annals-of-statistics/volume-38/issue-2/Nearly-unbiased-variable-selection-under-minimax-concave-penalty/10.1214/09-AOS729.short"
}

@ARTICLE{Daubechies2010-sj,
  title     = "Iteratively reweighted least squares minimization for sparse
               recovery",
  author    = "Daubechies, Ingrid and DeVore, Ronald and Fornasier, Massimo and
               Güntürk, C Si̇nan",
  journal   = "Commun. Pure Appl. Math.",
  publisher = "Wiley",
  volume    =  63,
  number    =  1,
  pages     = "1--38",
  month     =  jan,
  year      =  2010,
  url       = "http://dx.doi.org/10.1002/cpa.20303"
}

@ARTICLE{Daubechies2004-tx,
  title     = "An iterative thresholding algorithm for linear inverse problems
               with a sparsity constraint",
  author    = "Daubechies, I and Defrise, M and De Mol, C",
  journal   = "Commun. Pure Appl. Math.",
  publisher = "Wiley",
  volume    =  57,
  number    =  11,
  pages     = "1413--1457",
  month     =  nov,
  year      =  2004,
  url       = "http://dx.doi.org/10.1002/cpa.20042"
}

@ARTICLE{Donoho2009-jl,
  title     = "Message-passing algorithms for compressed sensing",
  author    = "Donoho, David L and Maleki, Arian and Montanari, Andrea",
  journal   = "Proc. Natl. Acad. Sci. U. S. A.",
  publisher = "Proceedings of the National Academy of Sciences",
  volume    =  106,
  number    =  45,
  pages     = "18914--18919",
  month     =  "10~" # nov,
  year      =  2009,
  url       = "http://dx.doi.org/10.1073/pnas.0909892106"
}

@ARTICLE{Krzakala2012-ot,
  title     = "Probabilistic reconstruction in compressed sensing: algorithms,
               phase diagrams, and threshold achieving matrices",
  author    = "Krzakala, Florent and Mézard, Marc and Sausset, Francois and Sun,
               Yifan and Zdeborová, Lenka",
  journal   = "J. Stat. Mech.",
  publisher = "IOP Publishing",
  volume    =  2012,
  number    =  08,
  pages     = "P08009",
  month     =  "17~" # aug,
  year      =  2012,
  url       = "https://iopscience.iop.org/article/10.1088/1742-5468/2012/08/P08009"
}

@ARTICLE{Candes2008-de,
  title   = "Enhancing Sparsity by Reweighted ℓ 1 Minimization",
  author  = "Candès, Emmanuel J and Wakin, Michael B and Boyd, Stephen P",
  journal = "J Fourier Anal Appl",
  volume  =  14,
  number  = "5-6",
  pages   = "877--905",
  series  = "Springer Texts in Statistics",
  month   =  "15~" # dec,
  year    =  2008,
  url     = "http://dx.doi.org/10.1007/s00041-008-9045-x"
}

@ARTICLE{Chartrand2007-so,
  title     = "Exact Reconstruction of Sparse Signals via Nonconvex Minimization",
  author    = "Chartrand, Rick",
  journal   = "IEEE Signal Processing Letters",
  publisher = "IEEE",
  volume    =  14,
  number    =  10,
  pages     = "707--710",
  month     =  oct,
  year      =  2007,
  url       = "http://dx.doi.org/10.1109/LSP.2007.898300"
}

@ARTICLE{Sakata2021-du,
  title     = "Perfect reconstruction of sparse signals with piecewise
               continuous nonconvex penalties and nonconvexity control",
  author    = "Sakata, Ayaka and Obuchi, Tomoyuki",
  journal   = "J. Stat. Mech.",
  publisher = "IOP Publishing",
  volume    =  2021,
  number    =  9,
  pages     =  093401,
  month     =  "1~" # sep,
  year      =  2021,
  url       = "http://dx.doi.org/10.1088/1742-5468/ac1403"
}

@ARTICLE{Prater-Bennette2022-rz,
  title     = "The proximity operator of the log-sum penalty",
  author    = "Prater-Bennette, Ashley and Shen, Lixin and Tripp, Erin E",
  journal   = "J. Sci. Comput.",
  publisher = "Springer Science and Business Media LLC",
  volume    =  93,
  number    =  3,
  pages     =  67,
  month     =  "25~" # dec,
  year      =  2022,
  url       = "http://dx.doi.org/10.1007/s10915-022-02021-4"
}

@INPROCEEDINGS{Chartrand2008-rl,
  title     = "Iteratively reweighted algorithms for compressive sensing",
  author    = "Chartrand, Rick and Yin, Wotao",
  booktitle = "2008 IEEE International Conference on Acoustics, Speech and
               Signal Processing",
  publisher = "IEEE",
  pages     = "3869--3872",
  month     =  "4~" # mar,
  year      =  2008,
  url       = "http://dx.doi.org/10.1109/ICASSP.2008.4518498"
}

@ARTICLE{Beck2009-kk,
  title     = "A fast iterative shrinkage-thresholding algorithm for linear
               inverse problems",
  author    = "Beck, Amir and Teboulle, Marc",
  journal   = "SIAM J. Imaging Sci.",
  publisher = "Society for Industrial \& Applied Mathematics (SIAM)",
  volume    =  2,
  number    =  1,
  pages     = "183--202",
  month     =  "4~" # jan,
  year      =  2009,
  url       = "http://dx.doi.org/10.1137/080716542"
}

@ARTICLE{Shen2013-ed,
  title     = "Exact reconstruction analysis of log-sum minimization for
               compressed sensing",
  author    = "Shen, Yanning and Fang, Jun and Li, Hongbin",
  journal   = "IEEE Signal Process. Lett.",
  publisher = "Institute of Electrical and Electronics Engineers (IEEE)",
  volume    =  20,
  number    =  12,
  pages     = "1223--1226",
  month     =  dec,
  year      =  2013,
  url       = "http://dx.doi.org/10.1109/LSP.2013.2285579"
}

@ARTICLE{Zou2006-qw,
  title     = "The adaptive lasso and its oracle properties",
  author    = "Zou, Hui",
  journal   = "J. Am. Stat. Assoc.",
  publisher = "Informa UK Limited",
  volume    =  101,
  number    =  476,
  pages     = "1418--1429",
  month     =  "1~" # dec,
  year      =  2006,
  url       = "http://dx.doi.org/10.1198/016214506000000735"
}

@ARTICLE{Fan2001-pc,
  title     = "Variable selection via nonconcave penalized likelihood and its
               oracle properties",
  author    = "Fan, Jianqing and Li, Runze",
  journal   = "J. Am. Stat. Assoc.",
  publisher = "Informa UK Limited",
  volume    =  96,
  number    =  456,
  pages     = "1348--1360",
  month     =  dec,
  year      =  2001,
  url       = "http://dx.doi.org/10.1198/016214501753382273"
}

@article{Candes2006Stable,
  title = {Stable Signal Recovery from Incomplete and Inaccurate Measurements},
  author = {Cand{\`e}s, Emmanuel J. and Romberg, Justin K. and Tao, Terence},
  year = 2006,
  journal = {Commun. Pure Appl. Math.},
  volume = {59},
  number = {8},
  pages = {1207--1223},
  issn = {1097-0312},
  doi = {10.1002/cpa.20124},
  url = {https://onlinelibrary.wiley.com/doi/abs/10.1002/cpa.20124},
  urldate = {2026-08-24},
  langid = {english},
}

@ARTICLE{Donoho2006-jx,
  title     = "Compressed sensing",
  author    = "Donoho, D L",
  journal   = "IEEE Trans. Inf. Theory",
  publisher = "Institute of Electrical and Electronics Engineers (IEEE)",
  volume    =  52,
  number    =  4,
  pages     = "1289--1306",
  month     =  apr,
  year      =  2006,
  url       = "http://dx.doi.org/10.1109/tit.2006.871582"
}

@ARTICLE{Boyd2010-hi,
  title     = "Distributed optimization and statistical learning via the
               alternating direction method of multipliers",
  author    = "Boyd, Stephen",
  journal   = "Found. Trends® Mach. Learn.",
  publisher = "Now Publishers",
  volume    =  3,
  number    =  1,
  pages     = "1--122",
  year      =  2010,
  url       = "http://dx.doi.org/10.1561/2200000016"
}

@ARTICLE{Carvalho2010-rz,
  title     = "The horseshoe estimator for sparse signals",
  author    = "Carvalho, C M and Polson, N G and Scott, J G",
  journal   = "Biometrika",
  publisher = "Oxford University Press (OUP)",
  volume    =  97,
  number    =  2,
  pages     = "465--480",
  month     =  "1~" # jun,
  year      =  2010,
  url       = "http://dx.doi.org/10.1093/biomet/asq017"
}

@ARTICLE{Nagano2024-ig,
  title   = "Effect of global shrinkage parameter of horseshoe prior in
             compressed sensing",
  author  = "Nagano, Yasushi and Hukushima, K",
  journal = "J. Stat. Mech.",
  year    =  2024,
  url     = "https://iopscience.iop.org/article/10.1088/1742-5468/ad3195"
}

@ARTICLE{Ganguli2010-ef,
  title     = "Statistical mechanics of compressed sensing",
  author    = "Ganguli, Surya and Sompolinsky, Haim",
  journal   = "Phys. Rev. Lett.",
  publisher = "American Physical Society (APS)",
  volume    =  104,
  number    =  18,
  pages     =  188701,
  month     =  "7~" # may,
  year      =  2010,
  url       = "http://dx.doi.org/10.1103/PhysRevLett.104.188701"
}

@ARTICLE{Kabashima2009-gx,
  title     = "A typical reconstruction limit for compressed sensing based
               {onLp}-norm minimization",
  author    = "Kabashima, Y and Wadayama, T and Tanaka, T",
  journal   = "J. Stat. Mech.",
  publisher = "IOP Publishing",
  volume    =  2009,
  number    =  09,
  pages     = "L09003",
  month     =  "21~" # sep,
  year      =  2009,
  url       = "http://dx.doi.org/10.1088/1742-5468/2009/09/L09003"
}

@ARTICLE{Nagano2023-kq,
  title     = "Phase transition in compressed sensing with horseshoe prior",
  author    = "Nagano, Yasushi and Hukushima, Koji",
  journal   = "Phys. Rev. E.",
  publisher = "American Physical Society",
  volume    =  107,
  number    = "3-1",
  pages     =  034126,
  month     =  "17~" # mar,
  year      =  2023,
  url       = "http://link.aps.org/pdf/10.1103/PhysRevE.107.034126"
}

\end{document}